\documentclass{aa}  

\usepackage{soul}
\usepackage{xcolor}
\usepackage{graphicx}
\usepackage{multirow}
\usepackage{hhline}
\usepackage{amsmath}

\usepackage{txfonts}
\usepackage{hyperref}
\hypersetup{
    linkcolor=blue,
    citecolor=blue,
    colorlinks=true,
    filecolor=blue,      
    urlcolor=blue
    }

\usepackage{float}

\begin{document} 

   \title{Can the splashback radius be an observable boundary of galaxy clusters?}

   \titlerunning{The splashback radius in regions of the Virgo replica}
   
   \author{Th\'eo Lebeau\inst{1} \thanks{Corresponding author: \href{mailto:theo.lebeau@universite-paris-saclay.fr}{theo.lebeau@universite-paris-saclay.fr}}, Stefano Ettori\inst{2,3,1}, Nabila Aghanim\inst{1}, Jenny G. Sorce\inst{4,1,5}}

   \authorrunning{Lebeau et al.}

   \institute{Université Paris-Saclay, CNRS, Institut d’Astrophysique Spatiale, 91405 Orsay, France
        \and
            INAF, Osservatorio di Astrofisica e Scienza dello Spazio, via Piero Gobetti 93/3, 40129 Bologna, Italy
        \and
          INFN, Sezione di Bologna, viale Berti Pichat 6/2, 40127 Bologna, Italy
        \and
             Université de Lille, CNRS, Centrale Lille, UMR 9189 CRIStAL, F-59000 Lille, France
        \and
            Leibniz-Institut für Astrophysik (AIP), An der Sternwarte 16, 14482 Potsdam, Germany}

   \date{Received 27 March 2024 / Accepted 24 May 2024}
 
  \abstract
  {The splashback radius was proposed as a physically motivated boundary of clusters as it sets the limit between the infalling and the orbitally dominated regions. However, galaxy clusters are complex objects connected to filaments of the cosmic web from which they accrete matter that disturbs them and modifies their morphology. In this context, estimating the splashback radius and the cluster boundary becomes challenging. In this work, we use a constrained hydrodynamical simulation replicating the Virgo cluster embedded in its large-scale structure to investigate the impact of its local environment on the splashback radius estimate. We identify the splashback radius from 3D radial profiles of dark matter density, gas density, and pressure in three regions representative of different dynamical states: accretion from spherical collapse, filaments, and matter outflow. We also identify the splashback radius from 2D-projected radial profiles of observation-like quantities: mass surface density, emission measure, and Compton-$y$. We show that the splashback radius mainly depends on the dynamics in each region and the physical processes traced by the different probes. We find multiple values for the splashback radius ranging from 3.3$\pm$0.2 to 5.5$\pm$0.3~Mpc. In particular, in the regions of collapsing and outflowing materials, the splashback radii estimated from gas density and pressure radial profiles overestimate that of the dark matter density profiles, which is considered the reference value given that the splashback radius was originally defined from dark matter simulations in pioneering works. Consequently, caution is required when using the splashback radius as a boundary of clusters, particularly in the case of highly disturbed clusters like Virgo. We conclude with a discussion of the detection of the splashback radius from pressure radial profiles, which could be more related to an accretion shock, and its detection from stacked radial profiles.}

   \keywords{Galaxies: clusters: individual: Virgo - Galaxies: clusters: intracluster medium - Methods: numerical}

   \maketitle

%

\section{Introduction}

Clusters of galaxies are tracers of the cosmic matter density and distribution as they are the most massive gravitationally bound structures in the Universe. They are thus used to put constraints on cosmological parameters via their number count \citep[e.g.][]{planck2014szclustercount,salvati2018constraints,aymerich2024cosmological} or the baryon fraction \citep[e.g.][]{wicker2023constraining}, which requires a precise and unbiased mass estimation and calibration (e.g. \citeauthor{salvati2019mass} \citeyear{salvati2019mass}, \citeauthor{gianfagna2021exploring} \citeyear{gianfagna2021exploring}, \citeauthor{lebeau2024mass} \citeyear{lebeau2024mass}, \citeauthor{aymerich2024cosmological} \citeyear{aymerich2024cosmological} and references therein). This, in turn, necessitates the definition of a cluster boundary in which the mass is estimated. Assuming that galaxy clusters formed via the spherical collapse of initial overdensities \citep{gunn1972infall}, we can estimate their mass in a sphere of a characteristic radius. Considering the self-similar evolution of clusters \citep{kaiser1986evolution}, characteristic radii are also useful for comparing clusters through normalised quantities.\\

In the past decade, a new physically motivated radius setting the transition from free-fall to orbital motion of dark matter (DM) particles in N-body simulations has been proposed \citep{diemer2014dependence,adhikari2014splashback,more2015splashback}: the splashback radius, $R_{\text{sp}}$. It has been studied in various N-body and hydrodynamical cosmological simulations using DM, baryons, and galaxy densities (e.g.\citeauthor{diemer2014dependence} \citeyear{diemer2014dependence}; \citeauthor{diemer2017splashback} \citeyear{diemer2017splashback}; \citeauthor{mansfield2017splashback} \citeyear{mansfield2017splashback}; \citeauthor{o2021splashback} \citeyear{o2021splashback}; \citeauthor{aung2021shock} \citeyear{aung2021shock}; \citeauthor{shin2023sets} \citeyear{shin2023sets}; \citeauthor{zhang2023effect} \citeyear{zhang2023effect}; \citeauthor{towler2024inferring} \citeyear{towler2024inferring}) or intracluster light \citep{deason2021stellar}. \citet{diemer2014dependence} and \citet{pizzardo2024splashback} also used galaxy radial velocities. In particular, \cite{pizzardo2024splashback} proposed to define $R_{\text{sp}}$ as the transition point from a concave to a convex shape — that is, the second derivative equal to zero — of the radial profile of galaxy radial velocities.\\

\cite{diemer2020universal} shows that using the splashback radius to define the halo mass provides mass functions that are more universal — in other words, more independent of redshift and cosmology as expected by theoretical models \citep[e.g.][]{press1974formation} — than mass functions based on the virial mass or spherical overdensities' masses. According to them, this confirms that $R_{\text{sp}}$ is a physically motivated definition of a DM halo boundary. However, $R_{\text{sp}}$, and thus $M_{\text{sp}}$, the mass enclosed in a sphere of this radius, was determined by following DM particles' trajectories in \cite{diemer2020universal} (see \citeauthor{mansfield2017splashback} \citeyear{mansfield2017splashback} for details). In contrast, in observations, we can only try to derive $R_{\text{sp}}$ from radial profiles of either the galaxy number density or radial velocities, weak lensing, pressure, and intracluster light. Given that galaxy clusters are connected to filaments of the cosmic web \citep[e.g.][]{cautun2014evolution,peebles2020large,gouin2021shape,gouin2022gas,gouin2023soft,galarraga2020populations,galarraga2021properties,galarraga2022relative}, defining their boundary in this complex environment is complicated. \\

Moreover, $R_{\text{sp}}$ is located in clusters' outskirts, which can be very challenging to observe due to the signal's faintness. Still, numerous tentative detections in various wavelengths were performed by stacking the observed signal of a sample of clusters to reach a sufficiently high signal-to-noise ratio. Then, $R_{\text{sp}}$ was identified as the minimum gradient of the radial profile (see more details in Sect. \ref{subsec:2.2}) in every observable. In the optical, galaxy number density  \citep[e.g.][]{more2016detection,chang2018splashback,shin2019measurement,adhikari2021probing,bianconi2021locuss,baxter2021shocks,rana2023erosita}, weak lensing \citep{contigiani2019weak,shin2021mass,fong2022first,giocoli2024amico}, and intracluster light \citep{gonzalez2021discovery} were used. However, in our case study we cannot stack over a cluster sample because we only consider one. Moreover, due to the small number of galaxies in a single cluster, the detection of $R_{\text{sp}}$ from the galaxy number density or radial velocity profile is hardly reliable. Finally, at sub-millimeter wavelengths, through the thermal Sunyaev-Zel'dovich \citep[tSZ,][]{sunyaev1972observations} effect, a pressure decrease in the intracluster medium (ICM) was detected and associated with the splashback radius \citep[e.g.][]{anbajagane2022shocks,anbajagane2023cosmological,rana2023erosita}.\\

In the present study, we investigate the sensitivity of $R_{\text{sp}}$ to the dynamics in cluster outskirts and its observability. To this end, we conduct a detailed study of a state-of-the-art constrained hydrodynamical simulation reproducing the Virgo cluster in its local environment \citep{sorce2021hydrodynamical}. This Virgo replica, like the real observed one (see \citeauthor{sorce2021hydrodynamical} \citeyear{sorce2021hydrodynamical}, \citeauthor{lebeau2024mass} \citeyear{lebeau2024mass} and references therein), is a highly disturbed cluster connected to multiple filaments; it is thus a particularly interesting test case. We first present the methodology in Sect.~\ref{sec:2}, including selecting regions of accreting, outflowing, and filament material to investigate the sensitivity of the splashback radius estimate to specific dynamics. We identify $R_{\text{sp}}$ either on DM or gas density 3D radial profiles as well as on pressure 3D radial profiles in Sect. \ref{sec:3}. Since we use a constrained replica of the Virgo cluster in its local environment (see \citeauthor{sorce2019virgo} \citeyear{sorce2019virgo} for the constraining method), we can produce projected maps along the line of sight from the Milky Way to Virgo in our simulation, which is consistent with Virgo's observations. We thus identify $R_{\text{sp}}$ on 2D-projected radial profiles extracted from maps of mass surface density (SD), X-rays' emission measure (EM), and the Compton-$y$ signal in Sect.~\ref{sec:4}. In Sect.~\ref{sec:5}, we summarise our results, draw perspectives about the identification of $R_{\text{sp}}$ in observations of Virgo or other clusters, discuss the use of pressure, electron density, and entropy as tracers of $R_{\text{sp}}$, and compare the impact of using median or mean profiles to identify $R_{\text{sp}}$. Finally, we present our conclusions in Sect.~\ref{sec:6}.

\begin{table*}
    \centering
    \caption{Virgo's characteristic radii (top) and splashback radii (bottom).}
    \renewcommand{\arraystretch}{1.1}
    \begin{tabular}{ c c c c }
    \hline \hline
    \multicolumn{4}{c}{Characteristic radius (Mpc)}\\
    \hline
    $R_{\text{500c}}=1.1$~Mpc & $R_{\text{200c}}=1.7~$Mpc & $R_{\text{vir}}=2.1~$Mpc & $R_{\text{200m}}=2.9~$Mpc \\
    \hline \hline \hline
    \multicolumn{4}{c}{Splashback radius (Mpc)}\\
    \hline
        3D & DM ($R_{\text{sp,DM}}$) & Gas ($R_{\text{sp,gas} }$)  & Pressure ($R_{\text{sp,press}}$)  \\
        \hline
         Collapsing Material & $3.4 \pm 0.2$ & $3.9 \pm 0.2$ & $4.3 \pm 0.3$ \\
         Outflowing Material & $4.3 \pm 0.3$ & $4.9 \pm 0.3$ & $5.5 \pm 0.3$ \\
         Filament Material & - &- & - \\
          Full Material & - & - & - \\
    \hline
    2D & Surface density ($R_{\text{sp,SD}}$)  & Emission measure ($R_{\text{sp,EM}}$)  & Compton-$y$ ($R_{\text{sp,y}}$)  \\
    \hline
    & $3.3 \pm 0.2$  & $3.9 \pm 0.2$ & $4.9 \pm 0.3$ \\
    \hline
    
    \end{tabular}
    \label{tab:radii}
    \tablefoot{ The splashback radii are identified using 3D radial profiles in the regions or using the full material and 2D-projected quantities. The value gives the centre of the bin of minimum gradient; the error bars are plus and minus one bin width at this radius.}
\end{table*}

\section{Methodology}
\label{sec:2}

In this section, we first introduce the simulated replica of Virgo, define characteristic radii of galaxy clusters, present the method adopted to compute 3D and 2D-projected radial profiles, and finally identify the regions our study focuses on.

\subsection{The Virgo replica simulation}

Recently, \cite{sorce2019virgo} produced replicas of Virgo within its cosmic environment by using constrained initial conditions of the local Universe (see \citeauthor{sorce2016cosmicflows} \citeyear{sorce2016cosmicflows} for details). This set of DM simulations is in remarkable agreement with observations. The most representative simulated halo was then re-used to produce a high-resolution hydrodynamical simulation of Virgo \citep{sorce2021hydrodynamical}, which also agrees well with observations. More details about the simulation and its agreement with observations can be found in \cite{sorce2021hydrodynamical} and \cite{lebeau2024mass}. \\

Like all the CLONE (Constrained LOcal and Nesting Environment) simulations \citep[e.g.][]{ocvirk2020cosmic,libeskind2020hestia,dolag2023simulating} using initial conditions of the local Universe from \cite{sorce2016cosmicflows} and \cite{sorce2018galaxyuphill}, the Virgo replica zoom-in simulation relies on the \cite{planckcosmoparam2014} cosmological parameters; namely, the total matter density, $\Omega_m=0.307$, dark energy density, $\Omega_{\Lambda}=0.693$, baryonic density, $\Omega_b=0.048$, amplitude of the matter power spectrum at 8~$\mathrm{Mpc~h^{-1}}$, $\sigma_8 = 0.829$, Hubble constant, $H_0=67.77~\mathrm{km~s^{-1}~Mpc^{-1}}$, and spectral index, $n_s=0.961$. This particular CLONE was produced using the {\tt RAMSES} \citep{teyssier2002cosmological} adaptive mesh refinement (AMR) code. The zoom region is contained in a 500~$\mathrm{Mpc~h^{-1}}$ local Universe box. It is a 30~Mpc diameter sphere with a resolution of $8192^3$ effective DM particles of mass $m_{\mathrm{DM}}=3\times10^7~\mathrm{M_\odot}$. The AMR grid has a finest cell size of 0.35~kpc. The simulation also contains sub-grid models for star formation, radiative gas cooling and heating, and kinetic feedback from the active galactic nucleus (AGN) and type II supernova (SN) similar to the Horizon-AGN implementation of \cite{dubois2014dancing,dubois2016horizon}. Furthermore, the AGN feedback model has been enhanced by orientating the jet according to the black hole (BH) spin (see \citeauthor{dubois2021introducing} \citeyear{dubois2021introducing} for details). \\

The properties of DM and gas cells were extracted using the {\tt rdramses} code\footnote{\url{https://github.com/florentrenaud/rdramses}.} first used in \cite{renaud2013sub}. The \cite{tweed2009building} halo finder has been used to identify the Virgo DM halo and its galaxies using the stars. In this work, we define the gas as the baryonic component in the simulation cells, in which hydrodynamics equations are solved following an Eulerian approach, and thus it does not include the stars. In the first approach, we did not apply any density or temperature selection; the gas thus includes its cold component from galaxies. All the gas cells and DM particles were used to compute the radial density profiles. On the contrary, to compute 3D pressure radial profile and produce Compton-$y$ maps, we selected cells of ionised gas — namely, cells with a temperature above $10^5$~K, which is the ionisation threshold — and removed the cells associated with the galaxies (see \citeauthor{lebeau2024mass} \citeyear{lebeau2024mass} for more details). To produce projected EM maps from the ICM gas, we only selected the cells with a temperature above $10^7$~K, given that this is approximately the minimum observable temperature of X-rays telescopes.

\subsection{Characteristic radii of galaxy clusters}
\label{subsec:2.2}

Assuming a spherical collapse model \citep{gunn1972infall}, we estimate the mass of a galaxy cluster within a sphere of a given radius defining its boundary. A first physically motivated boundary can be estimated using the virial theorem \citep{gunn1972infall,lacey1993merger,peebles2020large}. A cluster at dynamic equilibrium considered as virialised has a potential energy ($E_{\mathrm{p}}$) equal to the opposite of twice its kinetic energy ($E_{\mathrm{k}}$), $E_{\mathrm{p}} + 2E_{\mathrm{k}} = 0$. The distance at which this equation is valid is the virial radius ($R_{\text{vir}}$) and defines a boundary inside which the galaxy cluster matter is at dynamical equilibrium. Following this theorem, we can estimate cluster masses with galaxy velocities in optical galaxy surveys (e.g. \citeauthor{de1960apparent} \citeyear{de1960apparent} for the Virgo cluster) or in numerical simulations (see \citeauthor{vogelsberger2020cosmological} \citeyear{vogelsberger2020cosmological} for a review). For this Virgo replica simulation, we found $R_{\text{vir}}=2.1$~Mpc using the \cite{tweed2009building} halo finder .\\

However, $R_{\text{vir}}$ is not directly accessible when observing the ICM through the thermal Sunyaev-Zel'dovich effect or in the X-rays, since these observables trace the clusters' thermodynamic properties. Consequently, we usually estimate cluster masses using scaling relations \citep[][and references therein]{pratt2019galaxy} or the hydrostatic equilibrium \citep[e.g.][]{ettori2013mass} within radii enclosing a given overdensity. More precisely, we define a radius, $R_{\Delta \text{c}}$, at which the density is $\Delta$ times the critical density of the Universe:

\begin{equation}
    R_{\Delta \text{c}} = R ( \rho = \Delta \rho_{\text{c}} ) \quad \mathrm{with} \quad \rho_c=3H^2/8\pi G . 
\end{equation}

\noindent The mass enclosed in a sphere of this radius is thus 

\begin{equation}
    M_{\Delta \text{c}} = \Delta\rho_{\text{c}}\frac{4}{3}\pi R_{\Delta_c}^3
.\end{equation}

\noindent The most widely used characteristic radii are $R_{\text{500c}}$ and $R_{\text{200c}}$. For our Virgo replica, $R_{\text{500c}}=1.1$~Mpc and $R_{\text{200c}}=1.7~$Mpc.\\

Instead of using the critical density of the Universe, we can also use the mean matter density, $\rho_{\text{m}} = \rho_{\text{c}} \Omega_{\text{m}}$, and define $R_{\Delta \text{m}}$. Usually, we use $R_{\text{200m}}$ because it has been shown \citep[e.g.][]{kravtsov2012formation} that the ratio of the density at $R_{\text{vir}}$, $\rho_{\text{vir}}$, over $\rho_{\text{m}}$ is $\sim$200 ($\sim$178 in an Einstein-de Sitter Universe, see \citeauthor{bryan1998statistical} \citeyear{bryan1998statistical} for other cosmologies). To compare with other works, we normalise every radial profile by $R_{\text{200m}}$. For our Virgo replica, $R_{\text{200m}}=2.9$~Mpc. The values of these characteristic radii are summarised in Tab. \ref{tab:radii} for comparison with the $R_{\text{sp}}$ identified in this work.  \\

The focus of this study is the splashback radius, $R_{\text{sp}}$, introduced by works on N-body simulations \citep{adhikari2014splashback, diemer2014dependence,more2015splashback}. It was initially defined as the apocentre of the first orbit of DM particles free-falling onto a massive DM halo. It is a dynamic definition of a galaxy cluster since it sets the boundary between the region of infalling motion on the cluster and the region in which orbital motion dominates. Infalling particles will pile up at this radius as they inverse their radial velocity, so the particle density will strongly decrease beyond this characteristic radius. It can thus be identified as the steepest slope in the radial density profile. In other words, it is the minimum of the gradient of the radial density profile:

\begin{equation}
    R_{\text{sp}} = R\left[\text{Min}\left(\frac{\text{dlog}(\rho)}{\text{dlog}(r)}\right)\right],
\end{equation}

\noindent with Min for the minimum and $\rho$ the density of either DM, gas, or their sum. \\

\begin{figure*}
    \centering
    \includegraphics[trim=100 450 120 0, width=1\textwidth]{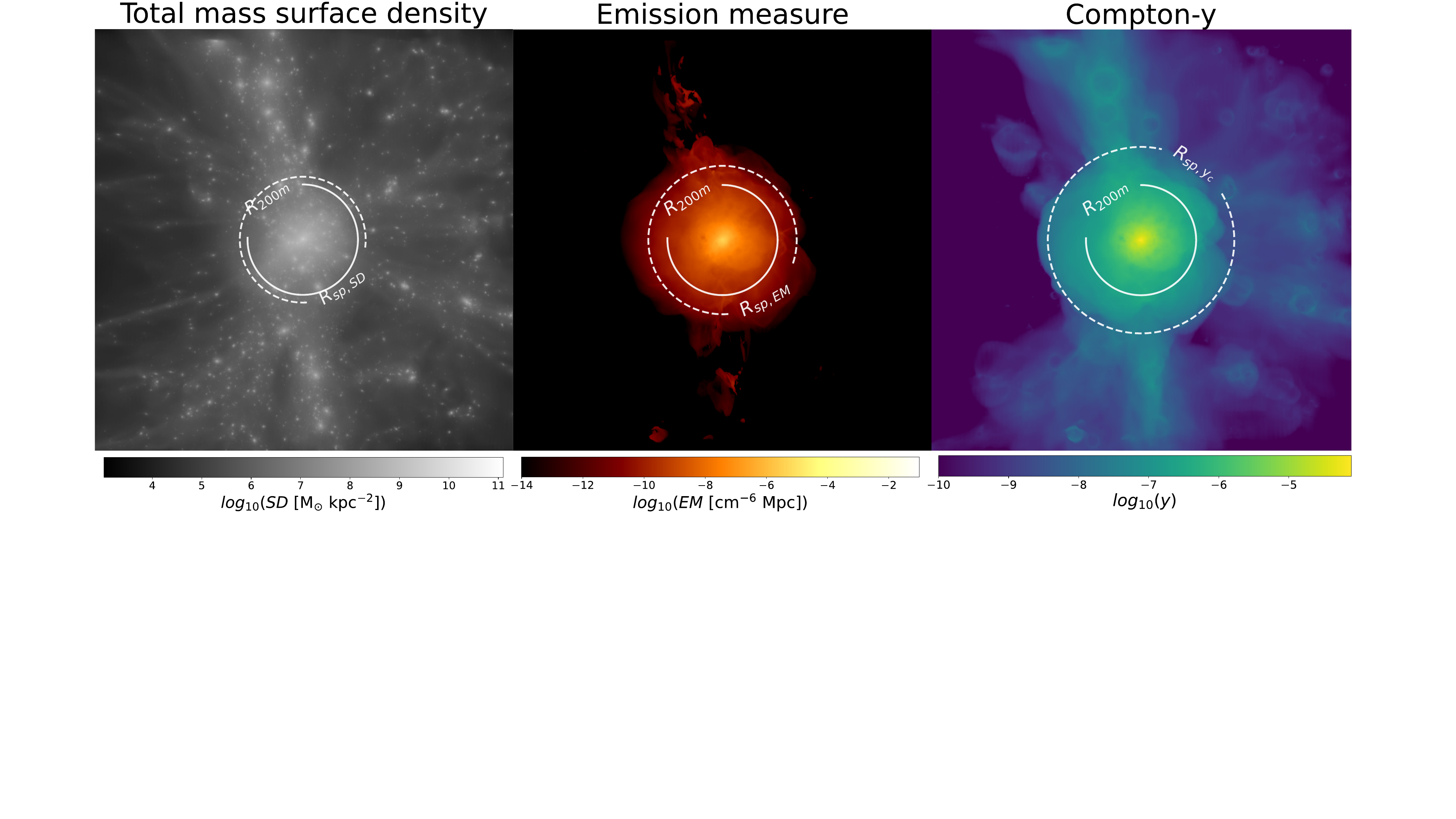}
    \caption{Projected maps of SD (left), EM (centre), and Compton-$y$ ($y$, right). The projection follows the direction between the Milky Way's assumed position and the Virgo cluster's centre. The maps are 22.122~Mpc wide and contain $15728^2$ pixels. The circles represent $R_{\text{200m}}$ (solid white) and $R_{\text{sp}}$ (dashed white) identified in each 2D-projected profile extracted from the maps. }
    \label{proj_maps}
\end{figure*}

\subsection{3D and 2D-projected radial profiles}

\begin{figure}         
            \centering
            \includegraphics[trim={250 50 600 0},clip, width=.49\textwidth]{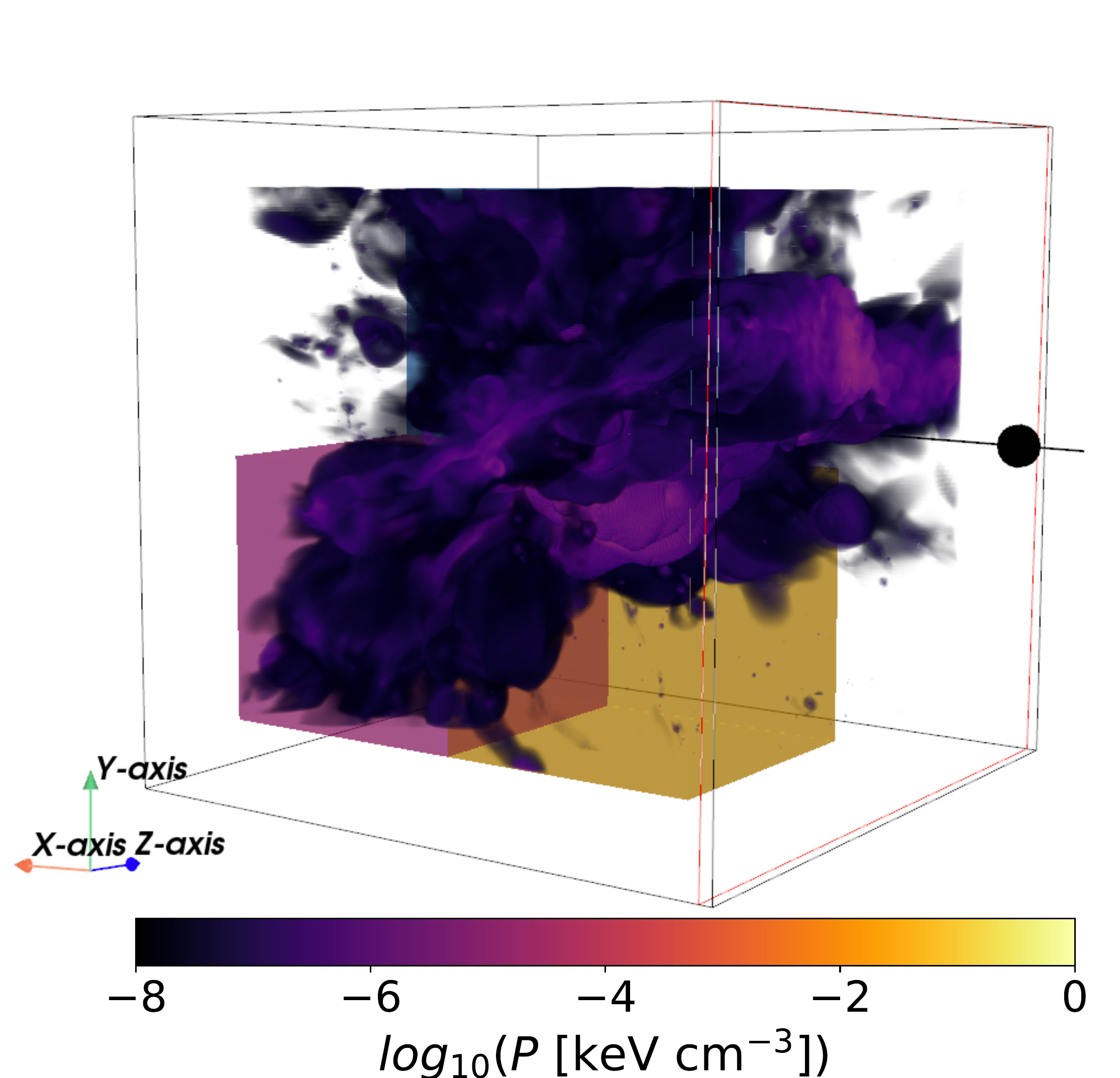}
            \hspace{0.1cm}
            \includegraphics[trim={250 50 600 0},clip, width=.49\textwidth]{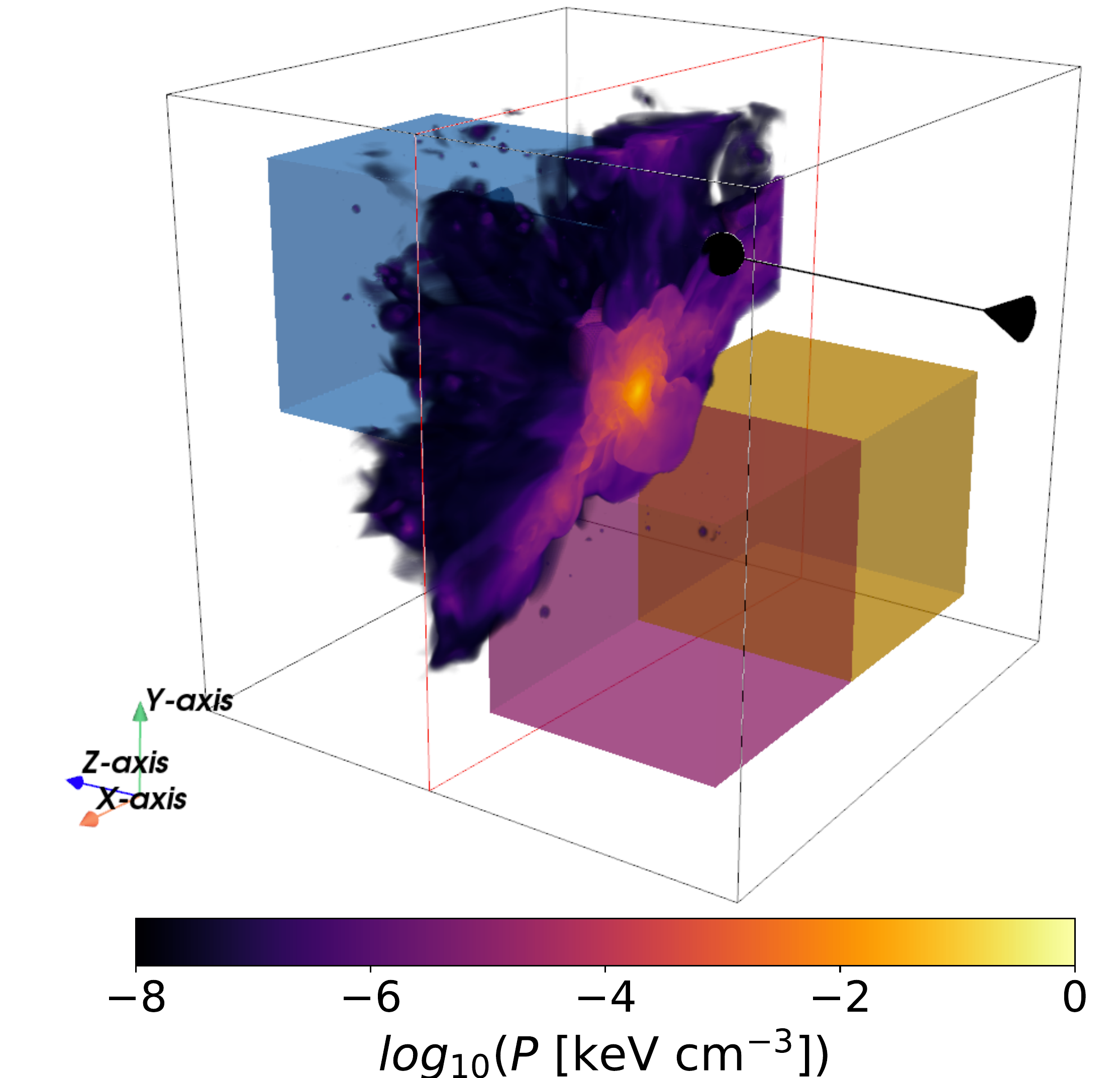}
            \caption{3D visualisation of the pressure in the Virgo replica simulation. The pressure is in $\mathrm{keV.cm^{-3}}$ and is shown on a log scale in the range of [-8,0]. The top panel shows the full box and the bottom panel shows a slice in the cluster's core along the z axis. The orange sub-box is the region of outflowing material, the blue sub-box is the region of collapsing material, and the pink sub-box is the region of filament material.}
            \label{3D_vis}
\end{figure}

In this work, we identify $R_{\text{sp}}$ as the radius of minimum gradient on radial profiles of DM, gas, and their sum, but also pressure, as other works \citep[e.g.][]{shi2016locations,towler2024inferring} relate it to $R_{\text{sp}}$. We also produce projected maps of observation-like quantities from which we extract 2D-projected radial profiles and once again identify $R_{\text{sp}}$ as the radius of minimum gradient, similar to \cite{towler2024inferring}. \\

We adapted the procedure presented in \cite{lebeau2024mass} to compute the radial profiles used in our analysis. We defined a binning of 0.025 on a log scale ranging from 200~kpc to the limit of the zoom-in region ($\sim$14.75~Mpc) minus the size of the biggest ($\sim$350~kpc) cells out of the zoom-in region. This avoided contamination from low-resolution cells at the borders of the zoom-in region, leading to non-physical profile shape. For 3D density profiles (DM, gas, and their sum in Figs. \ref{comp_bar_DM_sph_rel} and \ref{comp_bar_DM_fil+full}), we summed the masses in each spherical shell, and divided the summed mass by the volume of the spherical shell. For 3D gas properties (pressure, temperature, electron density, and entropy in Fig. \ref{thermo_profs} in this work), we computed their mass-weighted mean. The error bars are given by the mass-weighted standard deviation, and thus show the dispersion around the mean value. \\

We also computed observation-like quantities, from which we derived 2D-projected profiles and tried to identify $R_{\text{sp}}$. We built maps of SD akin to lensing signal, EM in X-rays, and integrated Compton-$y$ ($y$). The projected maps were built using the method detailed in \cite{lebeau2024mass}. They contain $15728^2$ pixels of 1.4065~kpc length and are 22.122~Mpc wide. We summed the contribution of all the cells along the line of sight in each pixel. In equations \ref{sigma_px},\ref{y_px} and \ref{em_px}, $m_{\text{cell,i}}$, $l_{\text{cell,i}}$, $P_{\text{e,cell,i}}$, $n_{\text{e,cell,i}}$, and $n_{\text{p,cell,i}}$ are the mass, length, electron pressure, electron density, and proton density of the $i$-th gas cell and $m_{\text{part,j}}$ is the mass of the $j$-th DM particle. \\

We built the SD map by summing the DM particles and gas cell masses in each pixel. For the gas cells contributing to more than one pixel, the mass attributed to each pixel is the cell's total mass divided by the number of occupied pixels. We then divided each pixel by its size; the map is in $\mathrm{M_{\odot}~kpc^{-2}}$. The method is summarised in the following equation giving the SD in a pixel, $\Sigma_{\text{px}}$:

\begin{equation}
    \Sigma_{\text{px}} = \left(\sum_i^{N_{\text{baryon\,cell\,in\,px}}}\frac{m_{\text{cell,i}}}{S_{\text{cell,i}}/S_{\text{px}}} + \sum_j^{N_{\text{DM\,particle\,in\,px}}}m_{\text{part,j}}\right)/S_{\text{px}}
    \label{sigma_px}
,\end{equation}

\noindent where $S_{\text{px}}$ is the pixel surface and $S_{\text{cell,i}}$ is the surface of the $i$-th cell. \\

The pressure in the ICM of galaxy clusters can be studied via the thermal Sunyaev-Zel'dovich effect (see \citeauthor{birkinshaw1999sunyaev} \citeyear{birkinshaw1999sunyaev} for a review) since it is proportional to the electron pressure integrated along the line of sight. To build the Compton-$y$ map, we summed the pressure in cells multiplied by their length in each pixel. The Compton-$y$ intensity in a pixel, $y_{\text{px}}$, is

\begin{equation}
    y_{\text{px}} = \frac{\sigma_Tk_{\text{b}}}{m_{\text{e}}c^2}\sum_i^{N_{\text{baryon\,cell\,in\,px}}}P_{\text{e,cell,i}}\;l_{\text{cell,i}}
    \label{y_px}
,\end{equation}

\noindent with $\sigma_T$ the Thomson cross section, $k_B$ the Boltzmann constant, $m_e$ the electron mass, and $c$ the speed of light. Compton-$y$ is unitless. \\

Finally, the density in the ICM is investigated through X-rays observations \citep{sarazin1986x}. In the X-rays soft band, the EM is the proton density times the electron density integrated along the line of sight. To build the map from the simulation box, we summed the product of the proton density and the electron density in the cells with a temperature higher than $10^7~\text{K}$ multiplied by their length in each pixel. The EM in a pixel, $EM_{\text{px}}$, in units of $\mathrm{cm^{-6}Mpc}$, is

\begin{equation}
    EM_{\text{px}} = \sum_i^{N_{\text{baryon\,cell\,in\,px}}}n_{\text{p,cell,i}}n_{\text{e,cell,i}}\;l_{\text{cell,i}}.
    \label{em_px}
\end{equation}

In Fig.~\ref{proj_maps}, we present the maps of SD (left), EM (centre), and Compton-$y$ (right). The color bar is on the log scale below each map. The solid white circle is $R_{\text{200m}}$; the dashed white circle is $R_{\text{sp}}$, identified on the 2D-projected radial profile extracted from the map (see Figs. \ref{proj_profs_SD}, \ref{proj_profs_EM} and \ref{proj_profs_y} in Sect. \ref{sec:4}). All the maps show the filamentary structure connected to Virgo: on the SD map we observe the DM halos' distribution, whereas on the EM and Compton-$y$ maps we see the diffuse gas, and more particularly tracers of its density and pressure, respectively. We also see an almost empty area on the left part of the Compton-$y$ map. \\

We then extracted 2D radial profiles (see Figs. \ref{proj_profs_SD}, \ref{proj_profs_EM} and \ref{proj_profs_y}) from the maps by computing the mean (similarly to pressure profiles, we discuss using the median instead in Sect. \ref{sec:5}) among the pixels in annuli, using the same radial binning as for the 3D spherical shells. The error bars are the standard deviation over the pixels in the circular annuli. We chose a projection following the direction between the centre of the large-scale simulation box, almost centred on the Milky Way, and the centre of Virgo; it is thus comparable to a real Virgo observation. For every radial profile presented in this paper (Figs.\ref{comp_bar_DM_sph_rel},\ref{comp_bar_DM_fil+full},\ref{thermo_profs},\ref{proj_profs_SD},\ref{proj_profs_EM} and \ref{proj_profs_y}), both axes are on a log scale and the radius in the abscissa was normalised by $R_{\text{200m}}=2.9~$Mpc introduced in Sect. \ref{subsec:2.2} (shown as a vertical dashed grey line) to compare with other studies.\\

\subsection{Selection of studied regions in Virgo}
\label{subsec:phase-space}

Clusters originally accreted matter via spherical collapse before relaxing and virialising. At present, they can still accrete matter from spherical collapse but also from cosmic filaments connected to them. These accretion regimes can take place at the same time in different regions of a single cluster. Hence, this work takes advantage of the zoom-in Virgo simulation embedded in its local environment to study the impact of the local environment, and so the dynamical state, on the estimation of the splashback radius in different regions. We selected three regions that are representative of distinct accretion regimes: accretion due to spherical collapse, accretion from filament, and matter outflow. They have the same volume and radial range, enabling direct one-to-one comparisons between regions. We defined these regions using both visual inspection and a radial distance-radial velocity phase-space diagram. \\

\begin{figure*}
    \centering
    \includegraphics[trim = 80 0 80 0, width=1\textwidth]{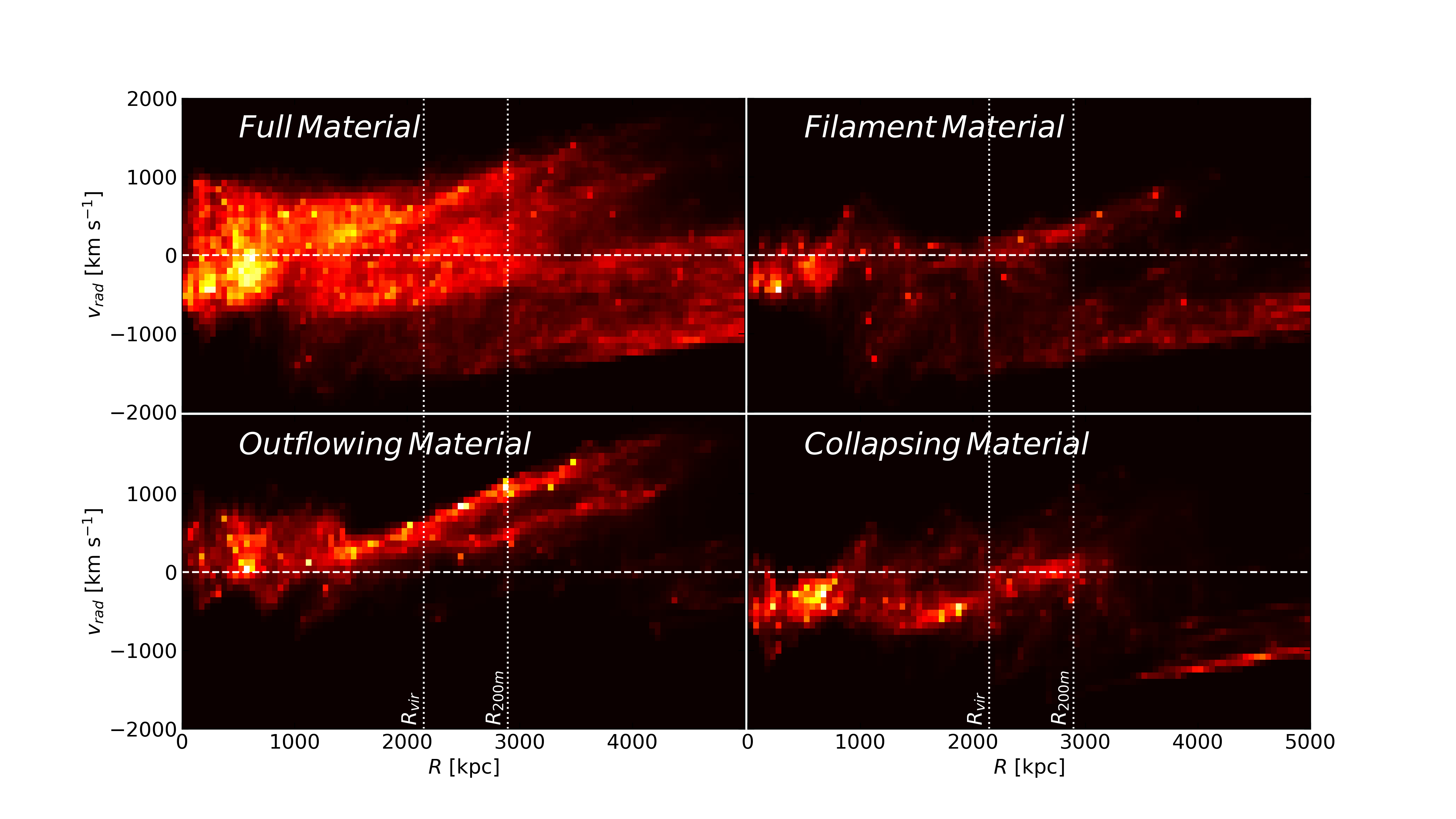} 
    \caption{Radial distance-radial velocity phase-space diagram of the ionised gas using the full material in the zoom-in box (top left), the region of filament material (top right), the region of outflowing material (bottom left) and the region of collapsing material (bottom right). In the region of filament material, most of the gas has a negative radial velocity of around -1000\,$\mathrm{km~s^{-1}}$, indicating that the gas is not slowed down while entering the ICM, which is typical of a filament. Then, in the region of collapsing material, most of the gas also has a negative radial velocity but this time of around -500\,$\mathrm{km~s^{-1}}$, meaning that in this region the gas is slowed down while being heated, which is the case for a spherical collapse. Finally, in the region of outflowing material, most of the gas has a positive radial velocity, and we thus have an outflow.}
    \label{phase_space}
\end{figure*}

First, the 3D visualisation\footnote{This clipping and slicing visualisation was made using the PyVista python library \citep{sullivan2019pyvista}} of the pressure in the zoom-in box used for the visual inspection is presented in Fig. \ref{3D_vis}. The top panel shows the full box, and the bottom panel shows a slice in the cluster core along the simulation's $z$ axis. 
Then, the radial distance-radial velocity phase-space diagram of the ionised gas, defined as gas cells not belonging to galaxies and with $T>10^5$~K, is presented in Fig. \ref{phase_space}. The radial velocity axis is in the range of [-2000,2000]~$\mathrm{km~s^{-1}}$ divided into 50 bins, and the radial distance is in the range of [0,5]~Mpc divided into 100 bins. All the ionised gas in the zoom-in region — Virgo and its neighbourhood — is shown in the top left panel. Throughout this paper, we refer to the full zoom-in region as `full material'. \\ 

In the pink box in Fig. \ref{3D_vis}, we see a large filament connected to Virgo that was already highlighted in \cite{lebeau2024mass}. When observing the phase-space diagram of this region, in the top right panel of Fig. \ref{phase_space}, we observe that the majority of the ionised gas, in the range of [1,5]~Mpc, has a high negative radial velocity, indicating that it is falling onto the cluster without being slowed down while entering the ICM. It confirms the presence of a filament shown in other works \citep[e.g.][]{gouin2022gas,vurm2023cosmic}. In the following, we call the matter in this region — that is, gas and DM — `filament material'.\\

Then, in the orange box in Fig. \ref{3D_vis}, we observe a minimal amount of pressure, and so matter, outside the cluster and an almost spherical cluster shape. In fact, its phase-space diagram, in the bottom left panel of Fig. \ref{phase_space}, shows that most of the ionised gas in the range of [1.5,4]~Mpc, has a positive radial velocity, indicating that it is moving away from the cluster. There is no accretion in this region as there is almost no ionised gas with a negative radial velocity outside the cluster; we thus call the matter in this region `outflowing material'.\\

Finally, in the blue box in Fig. \ref{3D_vis}, Virgo is also quite spherical in this region, but there is more infalling matter outside the cluster compared to the region of outflowing material. Its phase-space diagram, in the bottom right panel of Fig. \ref{phase_space}, shows that the ionised gas is falling on the cluster but with a much lower absolute radial velocity than for the filament material, particularly in the range of [1,2]~Mpc. It shows that it has been slowed down while entering the ICM. We consider that we are in the case of a spherical collapse; we thus call the matter in this region `collapsing material'.

\section{Splashback radius from 3D radial profiles}
\label{sec:3}

In this section, we identify $R_{\text{sp}}$ on 3D radial profiles. We first show the density profiles of DM, gas, and their sum in each region and compare the values found for these quantities. We then identify $R_{\text{sp}}$ on the pressure radial profile. The values of the identified splashback radii are summarised in Table \ref{tab:radii}.

\subsection{Splashback radius from 3D density profiles}

\begin{figure*}
            \sidecaption
            \includegraphics[width=12cm, trim = 50 50 90 20]{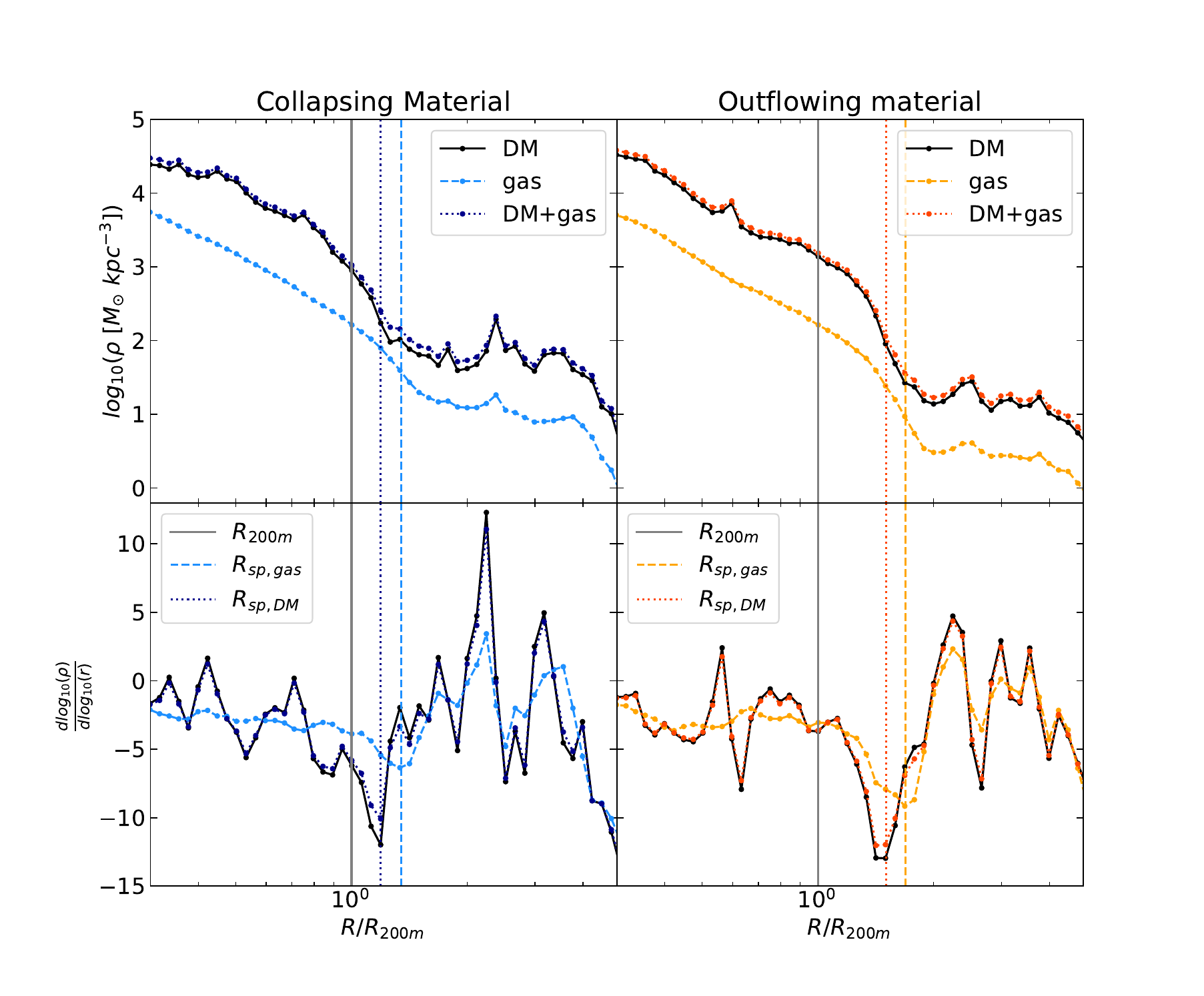}
            \caption{Comparison of radial density profiles for the region of collapsing material (left) and the region of outflowing material (right). For both sides, the top panel is the density profile and the bottom panel is the gradient. The solid black profile is the DM, the dashed coloured profile (blue for collapsing material and orange for outflowing material) is the gas, and its dotted dark counterpart is the sum of DM and gas. Vertical dashed lines are $R_{\text{200m}}$ (grey), $R_{\text{sp}}$ from gas (coloured), and DM (dark coloured). In each region, we observe that $R_{\text{sp,DM}}<R_{\text{sp,gas} }$. We also note that $R_{\text{sp}}$ is smaller for both DM and gas in the region of collapsing material compared to the region of outflowing material.}
            \label{comp_bar_DM_sph_rel}
\end{figure*}

\begin{figure*}
            \sidecaption
            \includegraphics[width=12cm,, trim = 50 50 90 20]{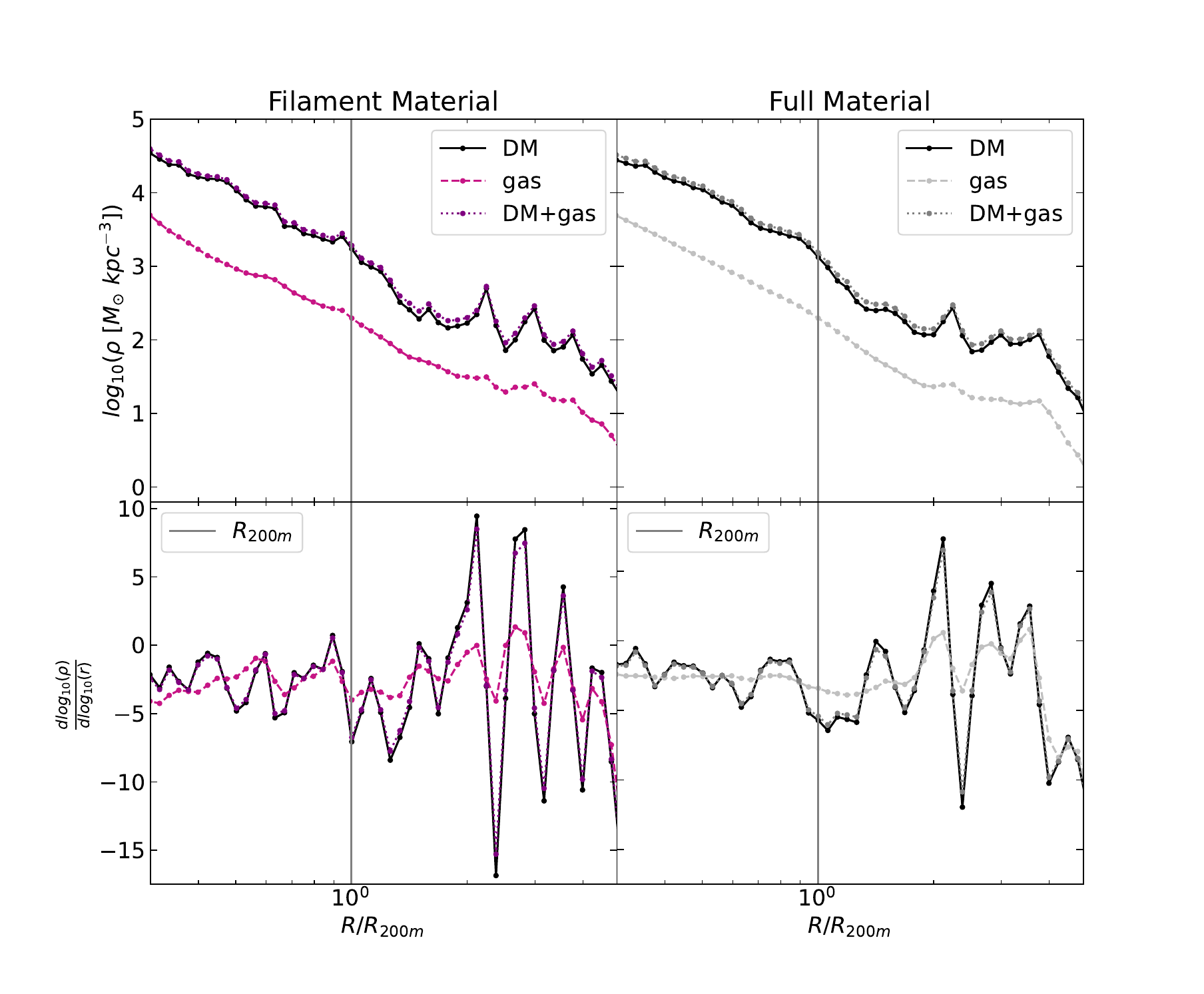}
            \caption{Comparison of radial density profiles for the region of filament material (left) and for the full material (right). For both sides, the top panel is the density profile and the bottom panel is the gradient. The solid black profile is the DM, the dashed coloured profile (pink for the region of filament material and grey for the full material) is the gas, and its dotted dark counterpart is the sum of DM and gas. The vertical dashed line is $R_{\text{200m}}$ (grey). $R_{\text{sp}}$ is not identified clearly in the region of filament material and when using the full material.}
            \label{comp_bar_DM_fil+full}
\end{figure*}

We present the density profiles of DM, gas, and their sum in the regions of collapsing material (Fig. \ref{comp_bar_DM_sph_rel}, left), outflowing material (Fig. \ref{comp_bar_DM_sph_rel}, right), filament material (Fig. \ref{comp_bar_DM_fil+full}, left), and full material (Fig. \ref{comp_bar_DM_fil+full}, right). In each figure, the radial profiles are in the top sub-panel and their gradient is in the bottom sub-panel; the profile displayed in black is the DM, and the profile displayed in light colours (respectively blue, orange, pink, and grey) is the gas.  Its dark-coloured counterpart is the sum of DM and gas.  \\

On the one hand, we clearly identify $R_{\text{sp}}$ either from DM and gas profiles in the regions of collapsing material (Fig. \ref{comp_bar_DM_sph_rel}, left) and outflowing material (Fig. \ref{comp_bar_DM_sph_rel}, right). For the former, we find $R_{\text{sp,DM}}=3.4\pm0.2~$Mpc (dashed vertical dark blue line) and $R_{\text{sp,gas} }=3.9\pm0.2~$Mpc (dashed vertical light blue line), and for the latter we find $R_{\text{sp,DM}}=4.3\pm0.3~$Mpc (dashed vertical dark orange line) and $R_{\text{sp,gas} }=4.9\pm0.3~$Mpc (dashed vertical light orange line). Moreover, we observe that the total matter density profiles are very similar to the DM profiles. It is expected given that DM largely dominates the mass budget in clusters; consequently, we find that $R_{\text{sp,tot}}=R_{\text{sp,DM}}$. \\

These identified $R_{\text{sp}}$ all have different values. First, when comparing $R_{\text{sp,DM}}$ and $R_{\text{sp,gas} }$ found in the region of collapsing (or outflowing) material, we observe that $R_{\text{sp,DM}}$ is about 0.5~Mpc (0.6~Mpc) smaller than $R_{\text{sp,gas} }$ – that is, a bin width of 3.5  (2) at this radius, which is quite significant. This can be explained by the fact that DM and gas  are not subject to the same physical processes; DM particles are collisionless and will thus not be slowed down while entering the cluster, whereas the gas has viscosity and will be slowed down and heated while encountering the ICM.\\ 

Then, when comparing $R_{\text{sp,DM}}$ (or $R_{\text{sp,gas} }$) between the regions of collapsing and outflowing materials, we observe a 0.9~Mpc (1~Mpc) difference, which is very significant given that it is three to four (three to five) bin widths at this radius. This reflects the dynamical state in each region; in the region of collapsing material, there is still free-fall matter accretion, so $R_{\text{sp}}$ is located at a smaller radius than in the region of outflowing material where the accretion rate is much lower. This was expected and is coherent with other studies (e.g. \citeauthor{diemer2014dependence} \citeyear{diemer2014dependence}, \citeauthor{o2021splashback} \citeyear{o2021splashback} and \citeauthor{towler2024inferring} \citeyear{towler2024inferring}) given that a high accretion rate induces a steeper potential well, leading to a smaller $R_{\text{sp}}$. According to \cite{towler2024inferring}, it can also be due to a high kinetic over thermal energy ratio; in our case, it might be a combination of both. \\

On the other hand, no clear feature is found in the gradient of either DM and gas density profiles in the region of filament material (Fig.\ref{comp_bar_DM_fil+full}, left) or with the full material (Fig.\ref{comp_bar_DM_fil+full}, right). There is an important minimum on one bin at $\sim 2.4~R_{\text{200m}}$ for the DM in the region of filament material, but this is certainly due to the presence of a massive galaxy or group of galaxies in the filament like all the other narrow peaks in the density profiles inducing a local maximum followed by a local minimum. Two other local minima are located close to $R_{\text{200m}}$ both for DM and gas  gradients, but they are quite weak compared to the $R_{\text{sp}}$ gradient in other regions, as is seen in Fig. \ref{comp_bar_DM_sph_rel}. Given that, in this region, the matter is funnelled in the cluster by the filament, we are not in the case of a spherical collapse; identifying $R_{\text{sp}}$ was thus not expected. The density profiles of the full material are quite similar to their counterparts in the region of filament material. This was expected, given that this region dominates the matter budget. We can observe minima for both the gas and DM gradient in the range of [2.6,4]~Mpc for DM and [2.9,4.3]~Mpc for gas.  However, the gradients in this range are not as low as the minimum gradients in the regions of outflowing and collapsing material. The contributions from all the regions are mixed, the identification of $R_{\text{sp}}$ is thus challenging. It only yields a mean value that does not encompass the full complexity of the cluster's dynamics. \\

\begin{figure*}
        \begin{minipage}[s]{1\textwidth}
            \centering
            \includegraphics[width=.44\textwidth]{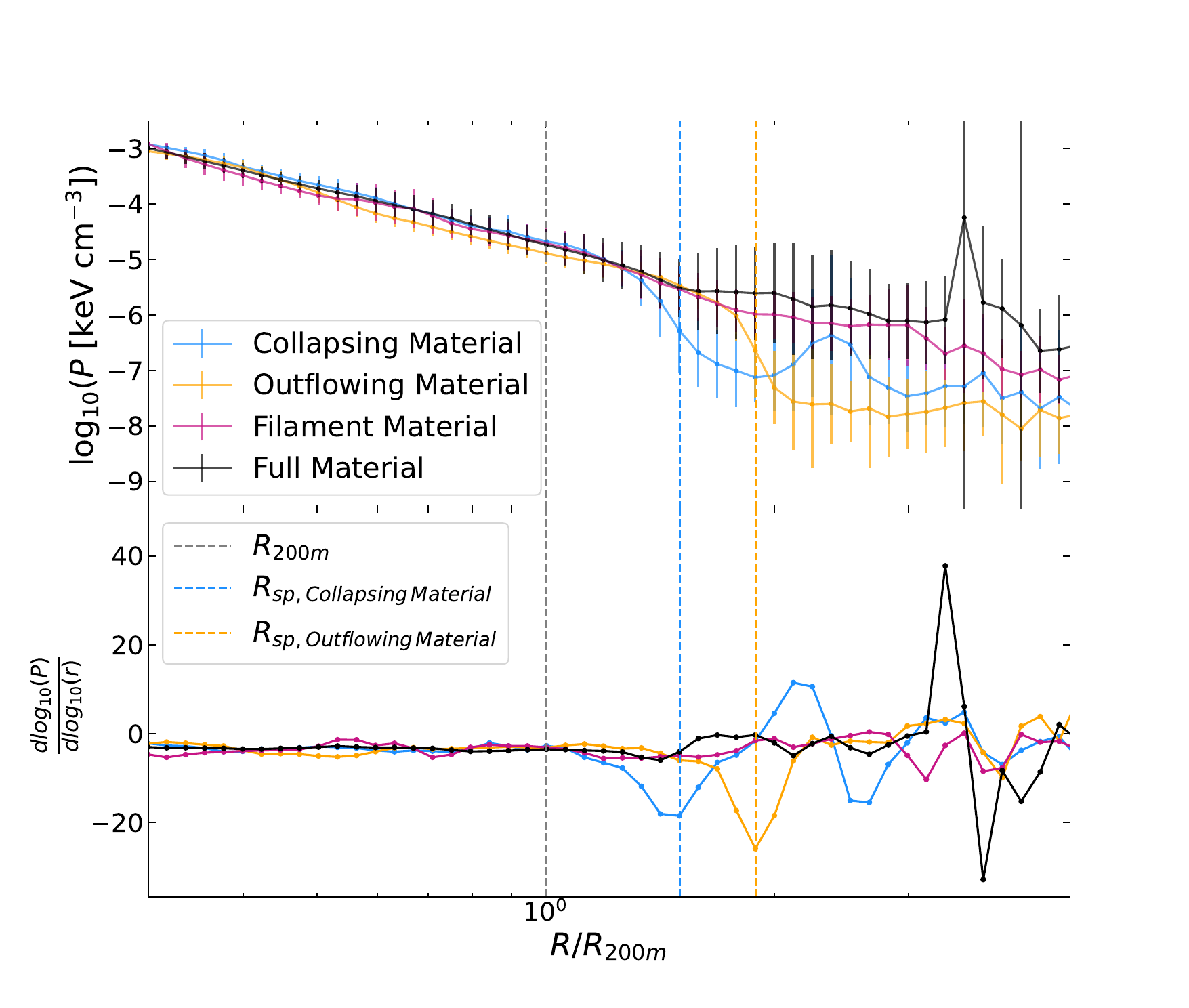}
            \hspace{0.1cm}
            \includegraphics[width=.54\textwidth]{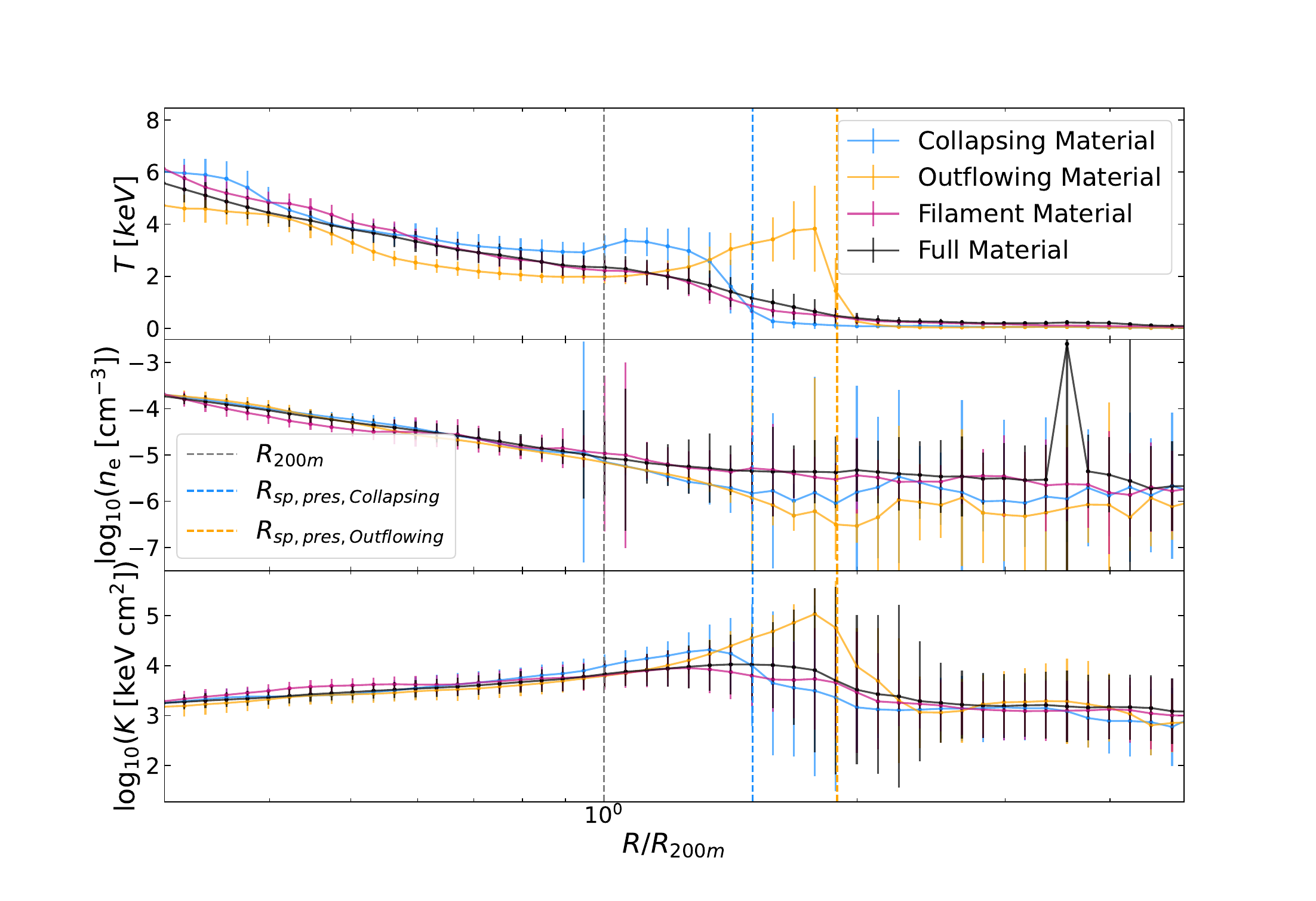}
            \caption{Left: Comparison of radial pressure profiles (top) and their gradient (bottom) in the selected regions and for the full material. Right: Comparison of  radial temperature (top), electron density (middle), and entropy (bottom) profiles also in the selected regions and for the full material. The colours stand for the same regions as for other figures: orange for the region of outflowing material, blue for the region of collapsing material, pink for the region of filament material, and black for the full material. $R_{\text{sp,press}}$ identified with the pressure profile with the outflowing material and the collapsing material are shown by dashed vertical orange and blue lines. $R_{\text{200m}}$ is shown in grey. We observe that $R_{\text{sp,gas} }<R_{\text{sp,press}}$ and that the minimum in the gradient corresponds to temperature and entropy peaks. Therefore, it might be an accretion shock rather than $R_{\text{sp}}$.}
            \label{thermo_profs}
       \end{minipage}
\end{figure*}

\subsection{Splashback radius from 3D pressure profiles}

In the left panel of Fig. \ref{thermo_profs}, we present the pressure radial profile (top sub-panel) and its gradient (bottom sub-panel). Similarly to Figs. \ref{comp_bar_DM_sph_rel} and \ref{comp_bar_DM_fil+full}, the profiles displayed in orange, blue, and pink are the regions of outflowing, collapsing, and filament material, respectively, and the profile displayed in black is the full material. For the full material, we observe a peak in the pressure profile at $\sim$3.5~$R_{\text{200m}}$ associated with a massive group of galaxies already identified in \cite{lebeau2024mass}. Apart from this feature, we find the same results as for the DM and gas density profiles (see Figs.\ref{comp_bar_DM_sph_rel} and \ref{comp_bar_DM_fil+full}): $R_{\text{sp,press}}$ is clearly identified in the regions of outflowing and collapsing material with more than one megaparsec difference whereas it is not in the region of filament material and for the full material. $R_{\text{sp,press}}=5.5\pm0.3~$Mpc in the region of outflowing material, and $R_{\text{sp,press}}=4.3\pm0.3~$Mpc in the region of collapsing material. They are highlighted by dashed vertical blue and orange lines. \\

These radii are larger than those extracted from the gas density profiles in those regions. However, $R_{\text{sp,press}}$ and $R_{\text{sp,gas} }$ in a given region are marginally compatible with their uncertainties. The pressure being the product of the electron density and the temperature means that this difference is due to the temperature. In fact, in the top right panel of Fig. \ref{thermo_profs}, we observe an extended temperature increase of up to a factor of 1.6 in the region of collapsing material and a factor of 6.7 in the region of outflowing material compared to the full material, which is compatible with both $R_{\text{sp,gas} }$ and $R_{\text{sp,press}}$. Multiple reasons could explain this temperature increase. First, we could have heating due to AGN feedback, coherently with \cite{towler2024inferring}, but it might not be powerful enough at these distances from the cluster's core to heat the ICM gas, and thus induce $R_{\text{sp,press}}$ to be hundreds of kpc further than $R_{\text{sp,gas} }$. $R_{\text{sp}}$ was originally defined from density profiles in N-body simulations. Thus, the most probable explanation is that we are not tracing $R_{\text{sp}}$ with pressure. We might rather be tracing accretion shocks, which is coherent with the strong temperature increase at these radii. We discuss this in Sect. \ref{sec:5}. Moreover, in the region of outflowing material, the shock in this region could be located at a larger radius in part due to the matter inflow. \\

In addition to the aforementioned shock fronts in the regions of outflowing and collapsing materials, we notice a temperature increase of up to a factor of 1.2 in the region of filament material but at a much smaller radius, identified in \cite{lebeau2024mass} at $\sim$850~kpc, which also indicates a shock front. This is due to the matter inflow in the filament penetrating deeply into the cluster without being slowed down and heated when encountering the ICM, as is shown in recent works (e.g. \citeauthor{gouin2022gas}  \citeyear{gouin2022gas}, \citeauthor{vurm2023cosmic} \citeyear{vurm2023cosmic}). None of these temperature increases are visible in the full material temperature profile because they are averaged over the overall mean temperature in each bin. \\

\begin{figure}[h!]

            \centering
            \includegraphics[trim=30 10 30 20, width=.41\textwidth]{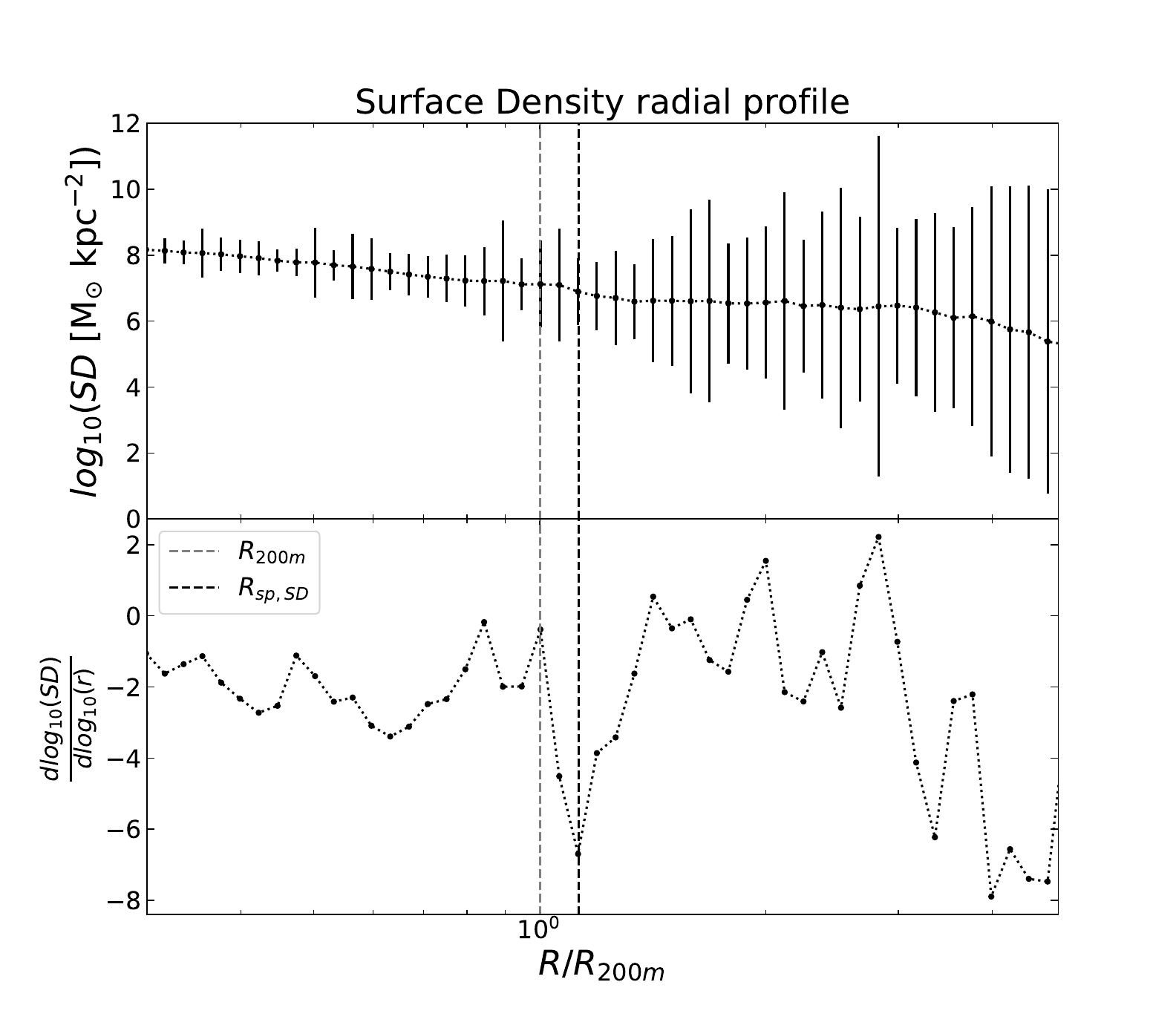}
            \caption{Projected radial profiles of SD. The profile is shown in the top sub-panel, and its gradient is shown in the bottom sub-panel. The dashed vertical lines represent $R_{\text{200m}}$ (grey) and $R_{\text{sp}}$ (black) identified in the 2D-projected profiles extracted from the SD map.}
            \label{proj_profs_SD}
    
\end{figure}

\begin{figure}
            \centering
            \includegraphics[trim=30 10 30 20, width=.41\textwidth]{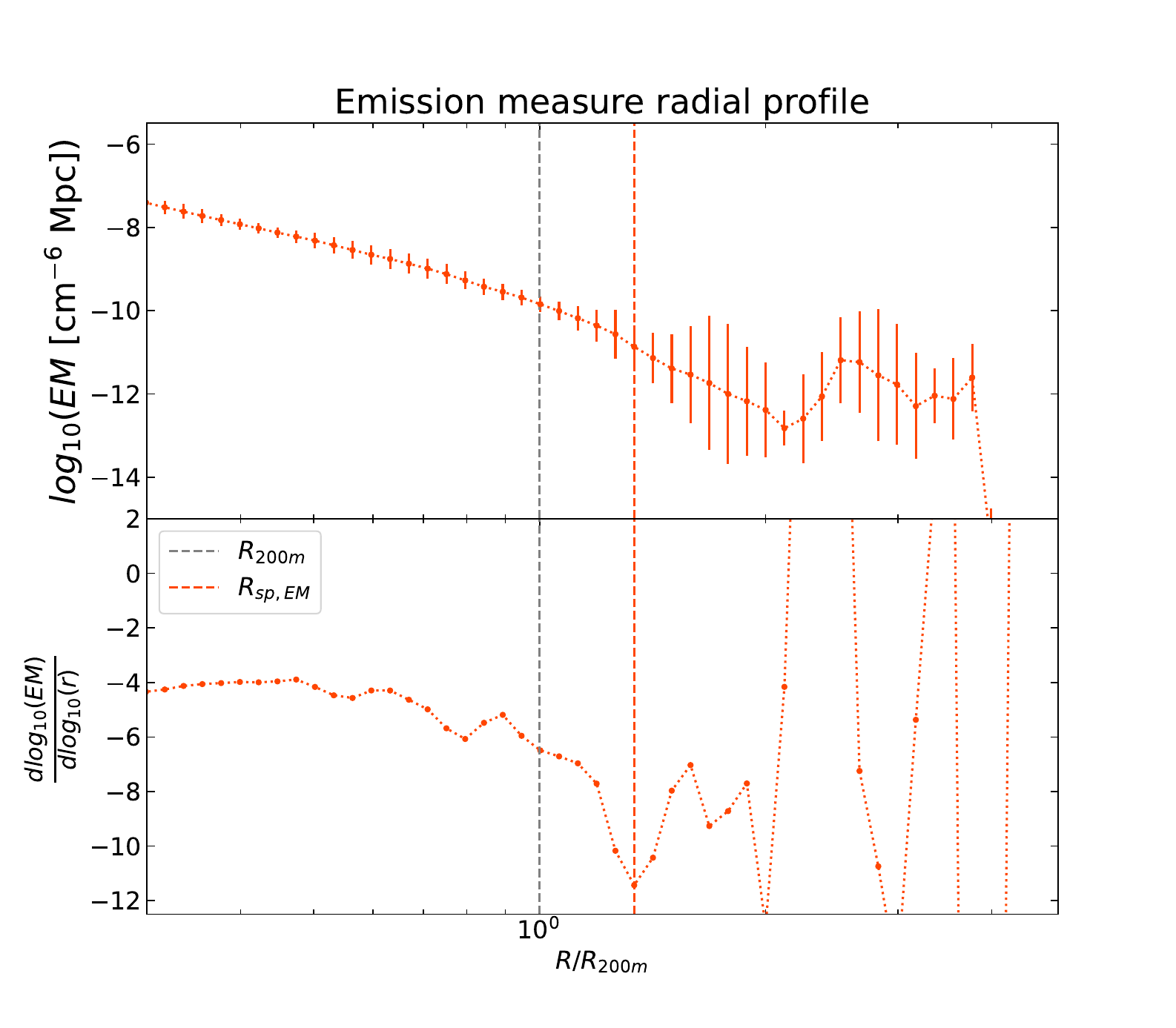}
            \caption{Projected radial profiles of EM. The profile is shown in the top sub-panel, and its gradient is shown in the bottom sub-panel. The dashed vertical lines represent $R_{\text{200m}}$ (grey) and $R_{\text{sp}}$ (orange) identified in the 2D-projected profiles extracted from the EM maps.}
            \label{proj_profs_EM}
            
\end{figure}

\begin{figure}
            \centering
            \includegraphics[trim=30 10 30 20, width=.41\textwidth]{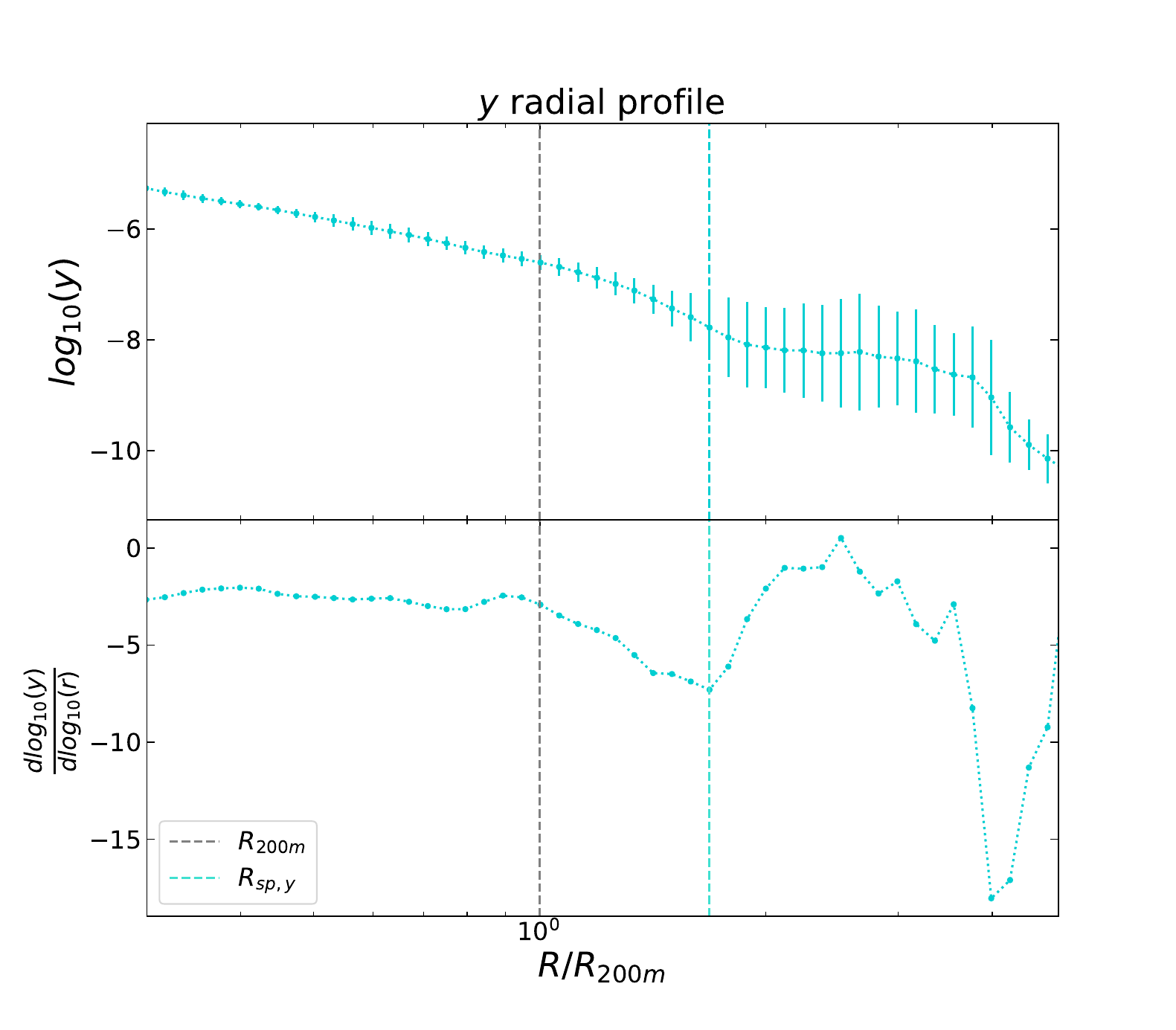}
            \caption{Projected radial profiles of Compton-$y$. The profile is shown in the top sub-panel, and its gradient is shown in the bottom sub-panel. The dashed vertical lines represent $R_{\text{200m}}$ (grey) and $R_{\text{sp}}$ (turquoise) identified in the 2D-projected profiles extracted from the Compton-$y$ maps.}
            \label{proj_profs_y}
            
\end{figure}

\section{Splashback radius from observation-like quantities}
\label{sec:4}

In this section, we use observation-like quantities projected along the Virgo-Milky-Way line of sight to identify $R_{\text{sp}}$. The maps and the method used to build them are presented in Sect. \ref{sec:2}. The 2D-projected radial profiles extracted from the maps (see again Sect. \ref{subsec:2.2} for details) are presented in Figs. \ref{proj_profs_SD}, \ref{proj_profs_EM}, and \ref{proj_profs_y}, showing respectively the SD, EM, and Compton-$y$. In each panel, $R_{\text{200m}}$ is represented by a dashed vertical grey line. We manage to identify $R_{\text{sp}}$ in each profile; the minimum in the gradient is clearly visible before a flattening of the profile induces an increase in the gradient. Their values can be found in Table \ref{tab:radii}, and their positions are highlighted on the profiles by a dashed vertical line, in black, orange, and turquoise, respectively.\\

When using the SD 2D-projected profile shown in Fig.\ref{proj_profs_SD}, we find that $R_{\text{sp,SD}}\sim3.3\pm0.2$~Mpc, in agreement with the value derived from the DM density profile in the region of collapsing material. This value is also in the range of the minimum gradient of the DM density profile of the full material (see right panel of Fig. \ref{comp_bar_DM_fil+full}). However, it differs from that derived from the gas density profile of the region of collapsing material. This was expected, given that DM dominates the mass budget in galaxy clusters, and so does the slope of the total mass density profile, as is shown in Figs. \ref{comp_bar_DM_sph_rel} and \ref{comp_bar_DM_fil+full}. Moreover, $R_{\text{sp,SD}}$ is not compatible with that found in the region of outflowing material, either using the DM, $R_{\text{sp,DM}}\sim4.3\pm0.3$~Mpc, or the gas, $R_{\text{sp,DM}}\sim4.9\pm0.3$~Mpc, density profile. The slope steepening of the radial profile that we associate with $R_{\text{sp}}$ extends over at least one order of magnitude in the DM profiles. Therefore, the steepening at the lowest radius among the regions determines that of the SD radial profile; it explains the agreement with the value found in the region of collapsing material. We can conclude that, in our case and in agreement with other works \citep[e.g.][]{towler2024inferring}, when using SD map to identify $R_{\text{sp}}$, we trace the DM dynamics in the region of collapsing material. \\

The EM traces the gas density as it is the integral of the electron density times the proton density. We find that $R_{\text{sp,EM}}\sim3.9~\pm0.2$~Mpc (see Fig. \ref{proj_profs_EM}), which is, this time, in agreement with the value derived from the gas density profile in the region of collapsing material. This was also expected, given that EM traces the gas density. Once again, the steepening at the lowest radius among the regions, the region of collapsing material in our case, determines that of the EM radial profile. The EM is thus a tracer of the gas dynamics in the region of collapsing material. \\ 

Finally, the thermal Sunyaev-Zel'dovich effect traces the pressure distribution; we find that $R_{\text{sp,yc}}\sim4.9~\pm0.3$~Mpc in the Compton-$y$ 2D-projected radial profile, shown in Fig.\ref{proj_profs_y}, which is much more distant from the centre than its SD and EM counterparts. This is in between the values extracted from the pressure profiles in the regions of outflowing and collapsing material. We would expect that $R_{\text{sp}}$ extracted from this 2D-projected profile agrees with that extracted from the pressure profile in the region of collapsing material, similarly to EM and SD. The low-pressure area visible on the left part of the map, inducing a substantial Compton-$y$ intensity decrease, is located in the region of outflowing material and undoubtedly impacts $R_{\text{sp}}$ identification. The contribution of the regions of outflowing and collapsing material seems then to be more mixed in this 2D-projected Compton-$y$ profile than in the SD and EM profiles due to this almost-empty region; it is another projection effect. The Compton-$y$ map traces the projected pressure distribution and gives much larger $R_{\text{sp}}$ than the SD and EM maps. However, it is still compatible with $R_{\text{sp,gas} }$ and $R_{\text{sp,press}}$ identified in the region of outflowing material.

\section{Discussion}
\label{sec:5}

\begin{figure}
            \centering
            \includegraphics[trim = 0 40 0 0, width=.5\textwidth]{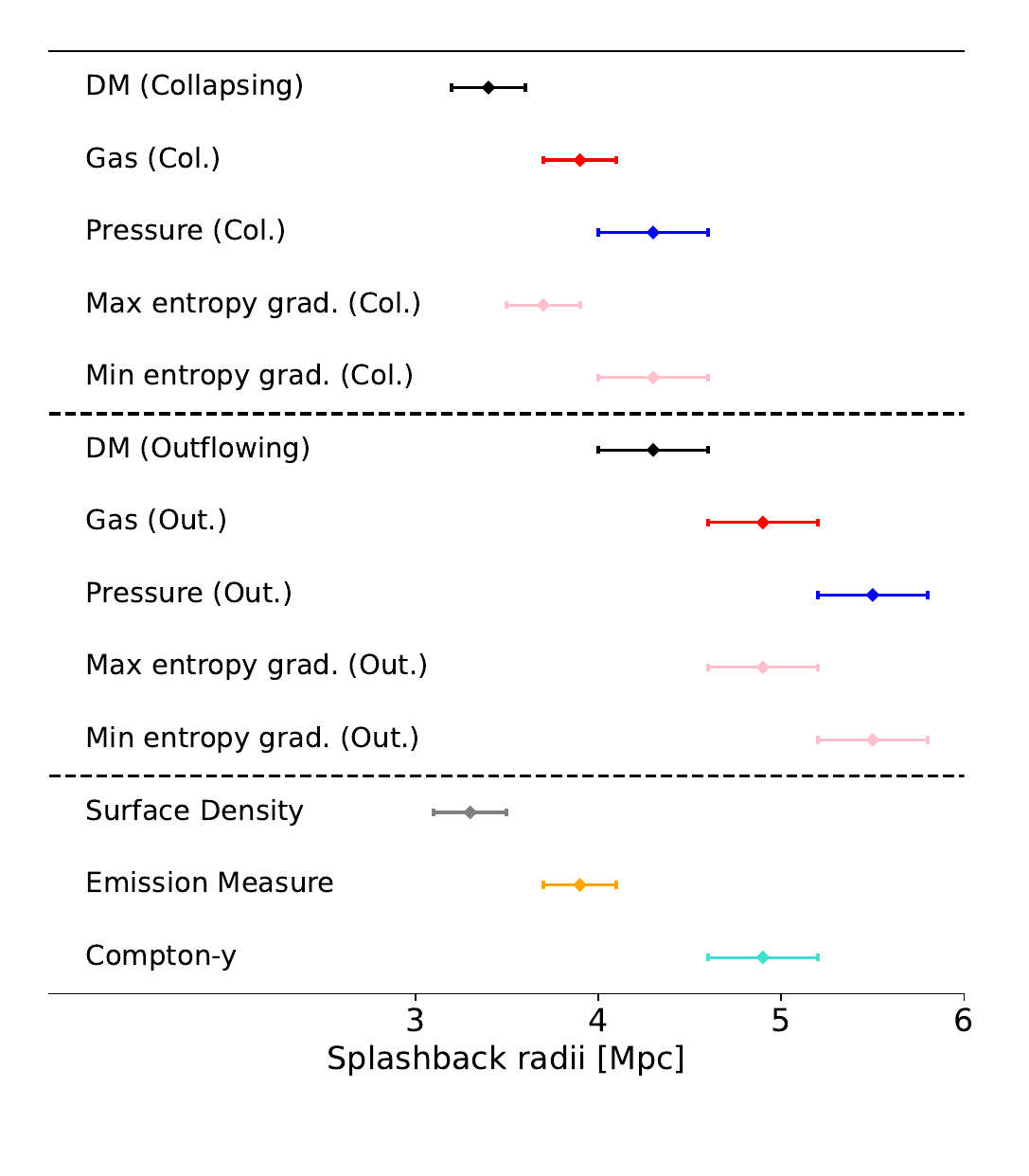}
            \caption{Comparison of the splashback radii obtained in this work. The values can be found in Tab. \ref{tab:radii}. In addition, the minimum and maximum gradients of entropy are shown in pink. In each region we observe that $R_{\text{sp,DM}}<R_{\text{sp,gas} }<R_{\text{sp,press}}$, and that $R_{\text{sp,SD}}$ is in good agreement with $R_{\text{sp,DM,Collapsing}}$ just like $R_{\text{sp,EM}}$ and $R_{\text{sp,gas,Collapsing}}$. We note that the minimum entropy gradient is in good agreement with $R_{\text{sp,press}}$, where the maximum entropy gradient corresponds well to $R_{\text{sp,gas} }$. } 
            \label{summary_fig_rsp}
\end{figure}

In this work, we used DM, gas, and total density as well as pressure radial profiles to identify $R_{\text{sp}}$. We also used projected observation-like quantities comparable to the aforementioned 3D profiles. We have found different values ranging from 3.3~Mpc to 5.5~Mpc; we compare them in Fig. \ref{summary_fig_rsp}. 
It shows that $R_{\text{sp,DM}}$, considered as the reference value since $R_{\text{sp}}$ was defined from DM simulations, has the smallest value in each region. $R_{\text{sp,gas} }$, and $R_{\text{sp,press}}$ even more, overestimates $R_{\text{sp}}$ compared to the $R_{\text{sp,DM}}$ reference value. 
However, as was discussed above, this is coherent since DM is collisionless whereas gas has viscosity, which leads to a larger gravitational potential well for the gas. In contrast, $R_{\text{sp,press}}$ might rather be an accretion shock, as we discuss in more detail below. Moreover, the radius found by each tracer is always at smaller radii in the region of collapsing material than in the region of outflowing material, which is consistent with other works \citep[e.g.][]{towler2024inferring} since the former region has a higher accretion rate, leading to a steeper gravitational potential well. Finally, $R_{\text{sp}}$ identified from 2D-projected radial profiles of SD, EM, and Compton-$y$ are in quite good agreement with their 3D counterpart (i.e. DM for SD, gas for EM, and pressure for Compton-$y$) in the region of collapsing material. Figure~\ref{summary_fig_rsp} thus highlights how much the identified $R_{\text{sp}}$ depends on the accretion regime in a given region and the probe used to identify it. \\

\subsection{Identification of $R_{sp}$ in observations}

On the observational side, given that Virgo is located in our direct vicinity, at approximatively 16 Mpc \citep{mei2007acs}, comprehensive studies of its galaxy population (e.g. the Next Generation Virgo Cluster Survey \citeauthor{ferrarese2016next} \citeyear{ferrarese2016next} or the Virgo Environmental Survey Tracing Ionised Gas Emission \citeauthor{boselli2018virgo} \citeyear{boselli2018virgo} surveys) have been carried out. We could thus derive its projected galaxy density or the galaxy radial velocities, but similarly to the present work, the limited number of detected galaxies in this mid-mass cluster, $M_{\text{vir}}\sim 6 \times 10^{14}~\mathrm{M_{\odot}}$ (\citeauthor{lebeau2024mass} \citeyear{lebeau2024mass} and references therein), could prevent the identification of $R_{\text{sp}}$ from galaxies.\\

Observing Virgo in the optical wavelength would be very challenging, then, given that it is very extended \citep[$R_{\text{vir}}$ angular size of 4°; see e.g. ][]{planck2016virgo}. Deriving its mass SD would therefore necessitate a dedicated observation program and enough luminous background sources to build a faithful lens model. To our knowledge, this has not been done so far; at best, galaxy-galaxy lensing could be performed in Virgo \citep{ferrarese2012next}. For other individual clusters, it might be possible to build a cluster lens model up to $R_{\text{sp}}$ provided there are sufficiently luminous and extended background sources.  \\

Observing Virgo's $R_{\text{sp}}$, or any other cluster, in the X-rays would be arduous too because the surface brightness is proportional to the squared density, leading to a very faint signal in the outskirts. As a matter of fact, \cite{urban2011x}, with XMM-Newton, and \cite{simionescu2015uniform}, with Suzaku, observed Virgo beyond $R_{\text{200c}}$ but only in particular arms using mosaics. The recently launched XRISM telescope \citep{xrism2020science,tashiro2022xrism} will not be able to study Virgo's outskirts due to its small collecting area; it will focus instead on the bright region Virgo where M87 is located. Recently, \cite{mccall2024srgerosita} took advantage of the SRG/eRosita all-sky survey to study the Virgo cluster up to 3~Mpc. It is unprecedented, but still does not reach the radii at which we identified $R_{\text{sp}}$ in this work, given that they are all beyond 3~Mpc (see Tab. \ref{tab:radii}). However, they conducted an azimuthal study similar to that performed both in observations \cite[e.g.][]{ettori1998rosat,vikhlinin1999outer,gouin2020probing} and simulations \cite{gouin2022gas}, and they probed an extended ICM emission and a cold front. With dedicated programs in the future, we might be able to conduct region studies at large radii like ours with observation data.\\

Finally, when considering the thermal Sunyaev-Zel'dovich effect, Virgo was only observed by the Planck telescope \citep{planck2016virgo}; the SPT and ACT ground telescopes cannot observe Virgo as it is close to the North Pole. Due to its proximity, Virgo has the largest integrated thermal Sunyaev-Zel'dovich flux \citep{taylor2003observing}, so it would be the best target for a first detection of $R_{\text{sp}}$ in a single cluster. Moreover, the thermal Sunyaev-Zel'dovich effect is the integrated pressure along the line of sight so it is proportional to electron density, whereas the surface brightness in the X-rays is proportional to the squared density, allowing one to study the ICM in the outskirts. However, \cite{planck2016virgo} could only study Virgo up to $\sim$1~Mpc, which would not be enough to identify $R_{\text{sp,press}}$. Future sub-millimetre experiments such as LiteBIRD \citep{litebird2023probing} might be able to observe clusters' outskirts up to $R_{\text{sp}}$. \\

\subsection{Pressure as a tracer of $R_{sp}$}

Similarly to what has been done with DM and gas density profiles, we identify $R_{\text{sp,press}}$ from the gas pressure radial profile in the regions of collapsing and outflowing materials, but at much larger radii. However, $R_{\text{sp}}$ is defined as a dynamical boundary; it is thus supposed to be identified from radial density profiles as it traces the underlying gravitational potential well-driving particles' velocity. The gas in the ICM experiences much more complex physical processes than DM, such as heating and turbulence due to AGN and SN feedback or shocks, particularly in the cluster's outskirts, where cosmic matter is accreted and shocked while entering the ICM. Therefore, when identifying the radius of the steepest slope on pressure radial profiles, we might identify an accretion shock rather than $R_{\text{sp}}$. \\

Some works based on simulations \citep[e.g.][]{shi2016locations,towler2024inferring} find a good agreement between the accretion shock radius, $R_{\text{sh}}$, and $R_{\text{sp}}$ whereas \cite{aung2021shock} find a ratio of $R_{\text{sh}}/R_{\text{sp}}\sim$1.89, where $R_{\text{sp}}$ is defined from DM particles' trajectories and $R_{\text{sh}}$ is defined as the radius of the minimum in the slope of the entropy profile. Moreover, \cite{anbajagane2022shocks} identify two minima in the stacked pressure profile of clusters observed with the South Pole Telescope (SPT; \citeauthor{plagge2010sunyaev} \citeyear{plagge2010sunyaev}). They associate the first one with a non-thermal equilibrium between ions and electrons occurring around shocks, as kinetic energy is better converted to thermal energy by ions than electrons, given that they are more massive. \\

However, this minimum cannot be observed and confirmed in our simulation since we assume local thermal equilibrium between ions and electrons. \cite{anbajagane2022shocks} identify the second minima as being related to an accretion shock front, for which they define a lower limit of $R_{\text{sh}}/R_{\text{sp}}\sim$2.16~$\pm$0.59. For simplicity, in the present work, we noted the minimum in the pressure gradient, $R_{\text{sp,press}}$, but in fact the ratio of $R_{\text{sp,press}}/R_{\text{sp,DM}}$ (similar to $R_{\text{sh}}/R_{\text{sp}}$ in \cite{aung2021shock}) is roughly 1.28 in the region of outflowing material and 1.26 in the region of collapsing material. It is a bit higher than one but much lower than the ratios found by \cite{aung2021shock} and \cite{anbajagane2022shocks}. Consequently, we are not able to state if $R_{\text{sh}}$ is distinguishable from $R_{\text{sp}}$ using the $R_{\text{sp,press}}/R_{\text{sp,DM}}$ ratio in our case study, although the temperature increase at these radii tends to motivate the presence of an accretion shock.\\

\subsection{Identification of $R_{\text{sp}}$ from electron density and entropy profiles}

Other thermodynamic quantities could be used as tracers of $R_{\text{sp}}$; for instance, it might in principle be possible to identify it from electron density profiles. In our case, we could not identify $R_{\text{sp}}$ in any of the electron density profiles (see middle right panel of Fig. \ref{thermo_profs}). Still, for the regions of outflowing and collapsing material, these radii seem to agree with their pressure counterparts since the steepening before the flattening of the profile occurs in the same range. Finally, the entropy, defined as $K=T/n_e^{2/3}$, is a tracer of the thermalisation state of the gas after being heated via shocks or adiabatic compression (see \citeauthor{tozzi2000detection} \citeyear{tozzi2000detection} for detection of entropy and \citeauthor{markevitch2007shocks} \citeyear{markevitch2007shocks} for a review on shocks). Both \cite{aung2021shock} and \cite{towler2024inferring} searched for the minimum gradient of the entropy profile to identify the accretion shock radius, as was discussed earlier. \cite{towler2024inferring} also find the maximum entropy gradient at approximately the splashback radius when defined from the gas density profile. This was more or less expected since when there is a minimum in the gas density gradient, there should be a corresponding maximum in the entropy gradient. In our work, in the regions of outflowing and collapsing material, the minimum gradient is at the same radius as $R_{\text{sp,press}}$ and the maximum is at approximately the same radius as $R_{\text{sp,gas} }$ (see the maximum and minimum entropy gradient lines displayed in pink in Fig. \ref{summary_fig_rsp}). This would tend to show that $R_{\text{sp,press}}$ is an accretion shock, even if, as was already discussed, we cannot confirm it from the $R_{\text{sp,press}}/R_{\text{sp,DM}}$ ratio. Anyway, the entropy seems to be a tracer of both $R_{\text{sp}}$ and $R_{\text{sh}}$. \\

\subsection{Comparison between the median and mean profiles}

As a baseline for our analysis, we chose the mean pressure and 2D-projected radial profiles. Computing the median profiles is a better tracer of an average quantity for log-normal distributions, which can be the case for the ICM properties in observations \citep[see e.g.][]{zhuravleva2013quantifying,eckert2015}. In our case, given that we selected the ionised gas cells (see Sect. \ref{sec:2}), that we have a very high number of cells per bin, and that the pressure is a quite smoothly distributed quantity, using the median instead of the mean gives quite similar results for 3D pressure profiles (see Fig. \ref{med_vs_mean_p} in Appendix \ref{mean_vs_med_appendix}) and for Compton-$y$ 2D-projected profiles (see Fig. \ref{stacked_profs_y} in Appendix \ref{stacking_appendix}). However, the SD intensity in pixels is not Gaussian-distributed. Indeed, there are important density contrasts due to sub-structures. Therefore, the median 2D-projected profile (see Fig. \ref{stacked_profs_SD} in Appendix \ref{stacking_appendix}) is quite different from its mean counterpart. In particular, there is an important dip within $R_{\text{200m}}$ causing a very important gradient. Though we doubt it can be associated with $R_{\text{sp}}$ at such a low radius, it emphasises the transition from the halo-dominated region to the background-dominated one. For EM, the median profile (see Fig.\ref{stacked_profs_EM} in Appendix \ref{stacking_appendix}) is consistent with the mean profile, including its gradient, up to $R_{\text{sp,EM}}$ but then gets steeper, so its minimum gradient is located a few bins further away. In this case, we are not able to conclude if this is due to the temperature cut, a projection effect, or if the median really identifies $R_{\text{sp,EM}}$ at a larger radius, moreover because the EM pixel distribution is not log-normal. In conclusion, even more caution is required when identifying $R_{\text{sp}}$ from EM and SD since the method used to compute the radial profile has a significant impact. \\

\subsection{Complementarity of this work with statistical studies}

We have shown that $R_{\text{sp}}$, more precisely the minimum in the gradient of a radial profile, seems to be more a tracer of the dynamical state of a cluster component (DM or gas) in a given region than an observable natural cluster boundary. However, our work suffers one major limitation, since it is a case study of a particularly disturbed cluster, while other cited works conducted statistical studies; our conclusions are thus not to be taken as universal. Yet this first case study of $R_{\text{sp}}$ in regions of a highly refined zoom-in simulation highlights how complex a single cluster can be. It also shows that when generalising over a cluster sample we mainly get a mean value averaging the signals in all the cluster environments. Specifically, when stacking over a large cluster sample, the resulting signal is averaged out, which smooths the outskirts, preventing one from identifying $R_{\text{sp}}$ (e.g. see the stacking of 100 random projections of Virgo in Figs. \ref{stacked_profs_SD}, \ref{stacked_profs_EM} and \ref{stacked_profs_y} in Appendix \ref{stacking_appendix}). Our case study is thus complementary to statistical studies that have tested the dependence of $R_{\text{sp}}$ on other parameters such as cluster mass, accretion rate, or redshift \citep{adhikari2014splashback,diemer2014dependence,more2015splashback,diemer2017splashback,mansfield2017splashback,o2021splashback}.

\section{Conclusion}
\label{sec:6}

In this work, we identified the splashback radius, $R_{\text{sp}}$ — the minimum in the gradient of the radial profile — using the radial density profiles of DM, gas, or their sum as well as pressure in regions of the simulated Virgo replica. These different regions, which contain outflowing, collapsing, and fragmented material, represent dynamical states occurring simultaneously in our Virgo case study. We then built projected maps of SD, EM, and Compton-$y$ along the Virgo-Milky Way line of sight in our constrained simulation of the local Universe. In addition, we discussed the use of different tracers to investigate $R_{\text{sp}}$; in particular, the pressure could be a tracer of an accretion shock rather than $R_{\text{sp}}$. We also discussed their potential detectability for a single cluster like Virgo, using median profiles instead of mean profiles to identify $R_{\text{sp}}$. We pointed out that our conclusions about this particularly disturbed cluster are not to be taken as universal, although this case study is complementary to statistical studies. \\

We have found that $R_{\text{sp}}$, when its identification is possible, varies by $\sim$1~Mpc between regions for a given quantity, which is about three to four bin widths at this radial distance from Virgo. We have also found that in the same region there is an almost 0.5~Mpc difference between the splashback radius extracted from the gas density profile and that extracted from the DM density profile, so two to three bin widths, which was expected, given that these components of galaxy clusters (i.e. gas  or DM) do not follow the same physical processes. Moreover, we notice that the contributions from all the regions are mixed when using the matter in the full zoom-in box to identify $R_{\text{sp}}$, which prevents us from identifying it clearly. \\

When extracting the splashback radius from 2D-projected profiles of SD, EM, and Compton-$y$, we have also found that each traces the physics processes of their underlying component. Moreover, the identified splashback radii are more in agreement with those found in the region of collapsing material than that of outflowing material. This is certainly because the slope steepening in this region determines that of the projected profile, given that it is located at a smaller radius than in the other regions, except for the Compton-$y$ profile, for which there is a consequent projection effect. \\

This study of regions in this simulated Virgo cluster shows the complex and various dynamics of clusters' outskirts and their impact on the splashback radius estimate. Caution is thus required when using the splashback radius as a natural boundary of clusters, especially for a complex cluster, as we have demonstrated for Virgo, and when stacking.

\begin{acknowledgements}
The authors acknowledge the Gauss Centre for Supercomputing e.V. (www.gauss-centre.eu) for providing computing time on the GCS Supercomputers SuperMUC at LRZ Munich. This work was supported by the grant agreements ANR-21-CE31-0019 / 490702358 from the French Agence Nationale de la Recherche / DFG for the LOCALIZATION project. 
This work has been supported as part of France 2030 program ANR-11-IDEX-0003. SE acknowledges the financial contribution from the contracts
Prin-MUR 2022 supported by Next Generation EU (n.20227RNLY3 {\it The concordance cosmological model: stress-tests with galaxy clusters}),
ASI-INAF Athena 2019-27-HH.0, ``Attivit\`a di Studio per la comunit\`a scientifica di Astrofisica delle Alte Energie e Fisica Astroparticellare'' (Accordo Attuativo ASI-INAF n. 2017-14-H.0),
from the European Union’s Horizon 2020 Programme under the AHEAD2020 project (grant agreement n. 871158), and the support by the Jean D'Alembert fellowship program.
The authors thank the IAS Cosmology team members for the discussions and comments. TL particularly thanks Stefano Gallo and Gaspard Aymerich for helpful discussions. 
The authors thank Florent Renaud for sharing the {\tt rdramses} {\tt RAMSES} data reduction code.
   
\end{acknowledgements}

\bibliographystyle{aa} 
\bibliography{bibliography}

\begin{thebibliography}{92}
\expandafter\ifx\csname natexlab\endcsname\relax\def\natexlab#1{#1}\fi

\bibitem[{Adhikari {et~al.}(2014)Adhikari, Dalal, \& Chamberlain}]{adhikari2014splashback}
Adhikari, S., Dalal, N., \& Chamberlain, R.~T. 2014, Journal of Cosmology and Astroparticle Physics, 2014, 019

\bibitem[{Adhikari {et~al.}(2021)Adhikari, Shin, Jain, Hilton, Baxter, Chang, Wechsler, Battaglia, Bond, Bocquet, {et~al.}}]{adhikari2021probing}
Adhikari, S., Shin, T.-h., Jain, B., {et~al.} 2021, The Astrophysical Journal, 923, 37

\bibitem[{Anbajagane {et~al.}(2023)Anbajagane, Chang, Baxter, Charney, Lokken, Aguena, Allam, Alves, Amon, An, {et~al.}}]{anbajagane2023cosmological}
Anbajagane, D., Chang, C., Baxter, E., {et~al.} 2023, Monthly Notices of the Royal Astronomical Society, stad3726

\bibitem[{Anbajagane {et~al.}(2022)Anbajagane, Chang, Jain, Adhikari, Baxter, Benson, Bleem, Bocquet, Calzadilla, Carlstrom, {et~al.}}]{anbajagane2022shocks}
Anbajagane, D., Chang, C., Jain, B., {et~al.} 2022, Monthly Notices of the Royal Astronomical Society, 514, 1645

\bibitem[{Aung {et~al.}(2021)Aung, Nagai, \& Lau}]{aung2021shock}
Aung, H., Nagai, D., \& Lau, E.~T. 2021, Monthly Notices of the Royal Astronomical Society, 508, 2071

\bibitem[{Aymerich {et~al.}(2024)Aymerich, Douspis, Pratt, Salvati, Soubri{\'e}, Andrade-Santos, Forman, Jones, Aghanim, Kraft, {et~al.}}]{aymerich2024cosmological}
Aymerich, G., Douspis, M., Pratt, G., {et~al.} 2024, arXiv preprint arXiv:2402.04006

\bibitem[{Baxter {et~al.}(2021)Baxter, Adhikari, Vega-Ferrero, Cui, Chang, Jain, \& Knebe}]{baxter2021shocks}
Baxter, E.~J., Adhikari, S., Vega-Ferrero, J., {et~al.} 2021, Monthly Notices of the Royal Astronomical Society, 508, 1777

\bibitem[{Bianconi {et~al.}(2021)Bianconi, Buscicchio, Smith, McGee, Haines, Finoguenov, \& Babul}]{bianconi2021locuss}
Bianconi, M., Buscicchio, R., Smith, G.~P., {et~al.} 2021, The Astrophysical Journal, 911, 136

\bibitem[{Birkinshaw(1999)}]{birkinshaw1999sunyaev}
Birkinshaw, M. 1999, Physics Reports, 310, 97

\bibitem[{Boselli {et~al.}(2018)Boselli, Fossati, Ferrarese, Boissier, Consolandi, Longobardi, Amram, Balogh, Barmby, Boquien, {et~al.}}]{boselli2018virgo}
Boselli, A., Fossati, M., Ferrarese, L., {et~al.} 2018, Astronomy \& Astrophysics, 614, A56

\bibitem[{Bryan \& Norman(1998)}]{bryan1998statistical}
Bryan, G.~L. \& Norman, M.~L. 1998, The Astrophysical Journal, 495, 80

\bibitem[{Cautun {et~al.}(2014)Cautun, Van De~Weygaert, Jones, \& Frenk}]{cautun2014evolution}
Cautun, M., Van De~Weygaert, R., Jones, B.~J., \& Frenk, C.~S. 2014, Monthly Notices of the Royal Astronomical Society, 441, 2923

\bibitem[{Chang {et~al.}(2018)Chang, Baxter, Jain, S{\'a}nchez, Adhikari, Varga, Fang, Rozo, Rykoff, Kravtsov, {et~al.}}]{chang2018splashback}
Chang, C., Baxter, E., Jain, B., {et~al.} 2018, The Astrophysical Journal, 864, 83

\bibitem[{Contigiani {et~al.}(2019)Contigiani, Hoekstra, \& Bah{\'e}}]{contigiani2019weak}
Contigiani, O., Hoekstra, H., \& Bah{\'e}, Y.~M. 2019, Monthly Notices of the Royal Astronomical Society, 485, 408

\bibitem[{de~Vaucouleurs(1960)}]{de1960apparent}
de~Vaucouleurs, G. 1960, The Astrophysical Journal, 131, 585

\bibitem[{Deason {et~al.}(2021)Deason, Oman, Fattahi, Schaller, Jauzac, Zhang, Montes, Bah{\'e}, Dalla~Vecchia, Kay, {et~al.}}]{deason2021stellar}
Deason, A.~J., Oman, K.~A., Fattahi, A., {et~al.} 2021, Monthly Notices of the Royal Astronomical Society, 500, 4181

\bibitem[{Diemer(2020)}]{diemer2020universal}
Diemer, B. 2020, The Astrophysical Journal, 903, 87

\bibitem[{Diemer \& Kravtsov(2014)}]{diemer2014dependence}
Diemer, B. \& Kravtsov, A.~V. 2014, The Astrophysical Journal, 789, 1

\bibitem[{Diemer {et~al.}(2017)Diemer, Mansfield, Kravtsov, \& More}]{diemer2017splashback}
Diemer, B., Mansfield, P., Kravtsov, A.~V., \& More, S. 2017, The Astrophysical Journal, 843, 140

\bibitem[{Dolag {et~al.}(2023)Dolag, Sorce, Pilipenko, Hern{\'a}ndez-Mart{\'\i}nez, Valentini, Gottl{\"o}ber, Aghanim, \& Khabibullin}]{dolag2023simulating}
Dolag, K., Sorce, J.~G., Pilipenko, S., {et~al.} 2023, Astronomy \& Astrophysics, 677, A169

\bibitem[{Dubois {et~al.}(2021)Dubois, Beckmann, Bournaud, Choi, Devriendt, Jackson, Kaviraj, Kimm, Kraljic, Laigle, {et~al.}}]{dubois2021introducing}
Dubois, Y., Beckmann, R., Bournaud, F., {et~al.} 2021, Astronomy \& Astrophysics, 651, A109

\bibitem[{Dubois {et~al.}(2016)Dubois, Peirani, Pichon, Devriendt, Gavazzi, Welker, \& Volonteri}]{dubois2016horizon}
Dubois, Y., Peirani, S., Pichon, C., {et~al.} 2016, Monthly Notices of the Royal Astronomical Society, 463, 3948

\bibitem[{Dubois {et~al.}(2014)Dubois, Pichon, Welker, Le~Borgne, Devriendt, Laigle, Codis, Pogosyan, Arnouts, Benabed, {et~al.}}]{dubois2014dancing}
Dubois, Y., Pichon, C., Welker, C., {et~al.} 2014, Monthly Notices of the Royal Astronomical Society, 444, 1453

\bibitem[{Eckert {et~al.}(2015)Eckert, Roncarelli, Ettori, Molendi, Vazza, Gastaldello, \& Rossetti}]{eckert2015}
Eckert, D., Roncarelli, M., Ettori, S., {et~al.} 2015, Monthly Notices of the Royal Astronomical Society, 447, 2198

\bibitem[{Ettori {et~al.}(2013)Ettori, Donnarumma, Pointecouteau, Reiprich, Giodini, Lovisari, \& Schmidt}]{ettori2013mass}
Ettori, S., Donnarumma, A., Pointecouteau, E., {et~al.} 2013, Space Science Reviews, 177, 119

\bibitem[{Ettori {et~al.}(1998)Ettori, Fabian, \& White}]{ettori1998rosat}
Ettori, S., Fabian, A., \& White, D. 1998, Monthly Notices of the Royal Astronomical Society, 300, 837

\bibitem[{Ferrarese {et~al.}(2012)Ferrarese, Cote, Cuillandre, Gwyn, Peng, MacArthur, Duc, Boselli, Mei, Erben, {et~al.}}]{ferrarese2012next}
Ferrarese, L., Cote, P., Cuillandre, J.-C., {et~al.} 2012, The Astrophysical Journal Supplement Series, 200, 4

\bibitem[{Ferrarese {et~al.}(2016)Ferrarese, C{\^o}t{\'e}, S{\'a}nchez-Janssen, Roediger, McConnachie, Durrell, MacArthur, Blakeslee, Duc, Boissier, {et~al.}}]{ferrarese2016next}
Ferrarese, L., C{\^o}t{\'e}, P., S{\'a}nchez-Janssen, R., {et~al.} 2016, The Astrophysical Journal, 824, 10

\bibitem[{Fong {et~al.}(2022)Fong, Han, Zhang, Yang, Gao, Wang, Li, Katsianis, \& Alonso}]{fong2022first}
Fong, M., Han, J., Zhang, J., {et~al.} 2022, Monthly Notices of the Royal Astronomical Society, 513, 4754

\bibitem[{Gal{\'a}rraga-Espinosa {et~al.}(2020)Gal{\'a}rraga-Espinosa, Aghanim, Langer, Gouin, \& Malavasi}]{galarraga2020populations}
Gal{\'a}rraga-Espinosa, D., Aghanim, N., Langer, M., Gouin, C., \& Malavasi, N. 2020, Astronomy \& Astrophysics, 641, A173

\bibitem[{Gal{\'a}rraga-Espinosa {et~al.}(2021)Gal{\'a}rraga-Espinosa, Aghanim, Langer, \& Tanimura}]{galarraga2021properties}
Gal{\'a}rraga-Espinosa, D., Aghanim, N., Langer, M., \& Tanimura, H. 2021, Astronomy \& Astrophysics, 649, A117

\bibitem[{Gal{\'a}rraga-Espinosa {et~al.}(2022)Gal{\'a}rraga-Espinosa, Langer, \& Aghanim}]{galarraga2022relative}
Gal{\'a}rraga-Espinosa, D., Langer, M., \& Aghanim, N. 2022, Astronomy \& Astrophysics, 661, A115

\bibitem[{Gianfagna {et~al.}(2021)Gianfagna, De~Petris, Yepes, De~Luca, Sembolini, Cui, Biffi, K{\'e}ruzor{\'e}, Mac{\'\i}as-P{\'e}rez, Mayet, {et~al.}}]{gianfagna2021exploring}
Gianfagna, G., De~Petris, M., Yepes, G., {et~al.} 2021, Monthly Notices of the Royal Astronomical Society, 502, 5115

\bibitem[{Giocoli {et~al.}(2024)Giocoli, Palmucci, Lesci, Moscardini, Despali, Marulli, Maturi, Radovich, Sereno, Bardelli, Castignani, Covone, Romanello, Roncarelli, \& Puddu}]{giocoli2024amico}
Giocoli, C., Palmucci, L., Lesci, G.~F., {et~al.} 2024 [\eprint[arXiv]{2402.06717}]

\bibitem[{Gonzalez {et~al.}(2021)Gonzalez, George, Connor, Deason, Donahue, Montes, Zabludoff, \& Zaritsky}]{gonzalez2021discovery}
Gonzalez, A.~H., George, T., Connor, T., {et~al.} 2021, Monthly Notices of the Royal Astronomical Society, 507, 963

\bibitem[{Gouin {et~al.}(2020)Gouin, Aghanim, Bonjean, \& Douspis}]{gouin2020probing}
Gouin, C., Aghanim, N., Bonjean, V., \& Douspis, M. 2020, Astronomy \& Astrophysics, 635, A195

\bibitem[{Gouin {et~al.}(2023)Gouin, Bonamente, Gal{\'a}rraga-Espinosa, Walker, \& Mirakhor}]{gouin2023soft}
Gouin, C., Bonamente, M., Gal{\'a}rraga-Espinosa, D., Walker, S., \& Mirakhor, M. 2023, Astronomy \& Astrophysics, 680, A94

\bibitem[{Gouin {et~al.}(2021)Gouin, Bonnaire, \& Aghanim}]{gouin2021shape}
Gouin, C., Bonnaire, T., \& Aghanim, N. 2021, Astronomy \& Astrophysics, 651, A56

\bibitem[{Gouin {et~al.}(2022)Gouin, Gallo, \& Aghanim}]{gouin2022gas}
Gouin, C., Gallo, S., \& Aghanim, N. 2022, Astronomy \& Astrophysics, 664, A198

\bibitem[{Gunn \& Gott~III(1972)}]{gunn1972infall}
Gunn, J.~E. \& Gott~III, J.~R. 1972, Astrophysical Journal, vol. 176, p. 1, 176, 1

\bibitem[{Kaiser(1986)}]{kaiser1986evolution}
Kaiser, N. 1986, Monthly Notices of the Royal Astronomical Society, 222, 323

\bibitem[{Kravtsov \& Borgani(2012)}]{kravtsov2012formation}
Kravtsov, A.~V. \& Borgani, S. 2012, Annual Review of Astronomy and Astrophysics, 50, 353

\bibitem[{Lacey \& Cole(1993)}]{lacey1993merger}
Lacey, C. \& Cole, S. 1993, Monthly Notices of the Royal Astronomical Society, 262, 627

\bibitem[{Lebeau {et~al.}(2024)Lebeau, Sorce, Aghanim, Hern{\'a}ndez-Mart{\'\i}nez, \& Dolag}]{lebeau2024mass}
Lebeau, T., Sorce, J.~G., Aghanim, N., Hern{\'a}ndez-Mart{\'\i}nez, E., \& Dolag, K. 2024, Astronomy \& Astrophysics, 682, A157

\bibitem[{Libeskind {et~al.}(2020)Libeskind, Carlesi, Grand, Khalatyan, Knebe, Pakmor, Pilipenko, Pawlowski, Sparre, Tempel, {et~al.}}]{libeskind2020hestia}
Libeskind, N.~I., Carlesi, E., Grand, R.~J., {et~al.} 2020, Monthly Notices of the Royal Astronomical Society, 498, 2968

\bibitem[{{LiteBIRD Collaboration}(2023)}]{litebird2023probing}
{LiteBIRD Collaboration}. 2023, Progress of Theoretical and Experimental Physics, 2023, 042F01

\bibitem[{Mansfield {et~al.}(2017)Mansfield, Kravtsov, \& Diemer}]{mansfield2017splashback}
Mansfield, P., Kravtsov, A.~V., \& Diemer, B. 2017, The Astrophysical Journal, 841, 34

\bibitem[{Markevitch \& Vikhlinin(2007)}]{markevitch2007shocks}
Markevitch, M. \& Vikhlinin, A. 2007, Physics Reports, 443, 1

\bibitem[{McCall {et~al.}(2024)McCall, Reiprich, Veronica, Pacaud, Sanders, Edler, Brüggen, Bulbul, de~Gasparin, Gatuzz, Liu, Merloni, Migkas, \& Zhang}]{mccall2024srgerosita}
McCall, H., Reiprich, T.~H., Veronica, A., {et~al.} 2024 [\eprint[arXiv]{2401.17296}]

\bibitem[{Mei {et~al.}(2007)Mei, Blakeslee, C{\^o}t{\'e}, Tonry, West, Ferrarese, Jord{\'a}n, Peng, Anthony, \& Merritt}]{mei2007acs}
Mei, S., Blakeslee, J.~P., C{\^o}t{\'e}, P., {et~al.} 2007, The Astrophysical Journal, 655, 144

\bibitem[{More {et~al.}(2015)More, Diemer, \& Kravtsov}]{more2015splashback}
More, S., Diemer, B., \& Kravtsov, A.~V. 2015, The Astrophysical Journal, 810, 36

\bibitem[{More {et~al.}(2016)More, Miyatake, Takada, Diemer, Kravtsov, Dalal, More, Murata, Mandelbaum, Rozo, {et~al.}}]{more2016detection}
More, S., Miyatake, H., Takada, M., {et~al.} 2016, The Astrophysical Journal, 825, 39

\bibitem[{Ocvirk {et~al.}(2020)Ocvirk, Aubert, Sorce, Shapiro, Deparis, Dawoodbhoy, Lewis, Teyssier, Yepes, Gottl{\"o}ber, {et~al.}}]{ocvirk2020cosmic}
Ocvirk, P., Aubert, D., Sorce, J.~G., {et~al.} 2020, Monthly Notices of the Royal Astronomical Society, 496, 4087

\bibitem[{O’Neil {et~al.}(2021)O’Neil, Barnes, Vogelsberger, \& Diemer}]{o2021splashback}
O’Neil, S., Barnes, D.~J., Vogelsberger, M., \& Diemer, B. 2021, Monthly Notices of the Royal Astronomical Society, 504, 4649

\bibitem[{Peebles(2020)}]{peebles2020large}
Peebles, P. J.~E. 2020, The large-scale structure of the universe, Vol.~98 (Princeton university press)

\bibitem[{Pizzardo {et~al.}(2024)Pizzardo, Geller, Kenyon, \& Damjanov}]{pizzardo2024splashback}
Pizzardo, M., Geller, M.~J., Kenyon, S.~J., \& Damjanov, I. 2024, Astronomy \& Astrophysics, 683, A82

\bibitem[{Plagge {et~al.}(2010)Plagge, Benson, Ade, Aird, Bleem, Carlstrom, Chang, Cho, Crawford, Crites, {et~al.}}]{plagge2010sunyaev}
Plagge, T., Benson, B., Ade, P.~A., {et~al.} 2010, The Astrophysical Journal, 716, 1118

\bibitem[{{Planck Collaboration}(2014{\natexlab{a}})}]{planckcosmoparam2014}
{Planck Collaboration}. 2014{\natexlab{a}}, A\&A, 571, A16

\bibitem[{{Planck Collaboration}(2014{\natexlab{b}})}]{planck2014szclustercount}
{Planck Collaboration}. 2014{\natexlab{b}}, Astronomy \& Astrophysics, 571, A20

\bibitem[{{Planck Collaboration}(2016)}]{planck2016virgo}
{Planck Collaboration}. 2016, Astronomy \& Astrophysics, 596, A101

\bibitem[{Pratt {et~al.}(2019)Pratt, Arnaud, Biviano, Eckert, Ettori, Nagai, Okabe, \& Reiprich}]{pratt2019galaxy}
Pratt, G., Arnaud, M., Biviano, A., {et~al.} 2019, Space Science Reviews, 215, 1

\bibitem[{Press \& Schechter(1974)}]{press1974formation}
Press, W.~H. \& Schechter, P. 1974, Astrophysical Journal, Vol. 187, pp. 425-438 (1974), 187, 425

\bibitem[{Rana {et~al.}(2023)Rana, More, Miyatake, Grandis, Klein, Bulbul, Chiu, Miyazaki, \& Bahcall}]{rana2023erosita}
Rana, D., More, S., Miyatake, H., {et~al.} 2023, Monthly Notices of the Royal Astronomical Society, 522, 4181

\bibitem[{Renaud {et~al.}(2013)Renaud, Bournaud, Emsellem, Elmegreen, Teyssier, Alves, Chapon, Combes, Dekel, Gabor, {et~al.}}]{renaud2013sub}
Renaud, F., Bournaud, F., Emsellem, E., {et~al.} 2013, Monthly Notices of the Royal Astronomical Society, 436, 1836

\bibitem[{Salvati {et~al.}(2018)Salvati, Douspis, \& Aghanim}]{salvati2018constraints}
Salvati, L., Douspis, M., \& Aghanim, N. 2018, Astronomy \& Astrophysics, 614, A13

\bibitem[{Salvati {et~al.}(2019)Salvati, Douspis, Ritz, Aghanim, \& Babul}]{salvati2019mass}
Salvati, L., Douspis, M., Ritz, A., Aghanim, N., \& Babul, A. 2019, Astronomy \& Astrophysics, 626, A27

\bibitem[{Sarazin(1986)}]{sarazin1986x}
Sarazin, C.~L. 1986, Reviews of Modern Physics, 58, 1

\bibitem[{Shi(2016)}]{shi2016locations}
Shi, X. 2016, Monthly Notices of the Royal Astronomical Society, 461, 1804

\bibitem[{Shin {et~al.}(2019)Shin, Adhikari, Baxter, Chang, Jain, Battaglia, Bleem, Bocquet, DeRose, Gruen, {et~al.}}]{shin2019measurement}
Shin, T.-h., Adhikari, S., Baxter, E., {et~al.} 2019, Monthly Notices of the Royal Astronomical Society, 487, 2900

\bibitem[{Shin \& Diemer(2023)}]{shin2023sets}
Shin, T.-h. \& Diemer, B. 2023, Monthly Notices of the Royal Astronomical Society, 521, 5570

\bibitem[{Shin {et~al.}(2021)Shin, Jain, Adhikari, Baxter, Chang, Pandey, Salcedo, Weinberg, Amsellem, Battaglia, {et~al.}}]{shin2021mass}
Shin, T.-h., Jain, B., Adhikari, S., {et~al.} 2021, Monthly Notices of the Royal Astronomical Society, 507, 5758

\bibitem[{Simionescu {et~al.}(2015)Simionescu, Werner, Urban, Allen, Ichinohe, \& Zhuravleva}]{simionescu2015uniform}
Simionescu, A., Werner, N., Urban, O., {et~al.} 2015, The Astrophysical Journal Letters, 811, L25

\bibitem[{Sorce(2018)}]{sorce2018galaxyuphill}
Sorce, J.~G. 2018, Monthly Notices of the Royal Astronomical Society, 478, 5199

\bibitem[{Sorce {et~al.}(2019)Sorce, Blaizot, \& Dubois}]{sorce2019virgo}
Sorce, J.~G., Blaizot, J., \& Dubois, Y. 2019, Monthly Notices of the Royal Astronomical Society, 486, 3951

\bibitem[{Sorce {et~al.}(2021)Sorce, Dubois, Blaizot, McGee, Yepes, \& Knebe}]{sorce2021hydrodynamical}
Sorce, J.~G., Dubois, Y., Blaizot, J., {et~al.} 2021, Monthly Notices of the Royal Astronomical Society, 504, 2998

\bibitem[{Sorce {et~al.}(2016)Sorce, Gottl{\"o}ber, Yepes, Hoffman, Courtois, Steinmetz, Tully, Pomarede, \& Carlesi}]{sorce2016cosmicflows}
Sorce, J.~G., Gottl{\"o}ber, S., Yepes, G., {et~al.} 2016, Monthly Notices of the Royal Astronomical Society, 455, 2078

\bibitem[{Sullivan \& Kaszynski(2019)}]{sullivan2019pyvista}
Sullivan, B. \& Kaszynski, A. 2019, Journal of Open Source Software, 4, 1450

\bibitem[{Sunyaev \& Zeldovich(1972)}]{sunyaev1972observations}
Sunyaev, R. \& Zeldovich, Y.~B. 1972, Comments on Astrophysics and Space Physics, 4, 173

\bibitem[{Tashiro(2022)}]{tashiro2022xrism}
Tashiro, M.~S. 2022, International Journal of Modern Physics D, 31, 2230001

\bibitem[{Taylor {et~al.}(2003)Taylor, Moodley, \& Diego}]{taylor2003observing}
Taylor, J.~E., Moodley, K., \& Diego, J. 2003, Monthly Notices of the Royal Astronomical Society, 345, 1127

\bibitem[{Teyssier(2002)}]{teyssier2002cosmological}
Teyssier, R. 2002, Astronomy \& Astrophysics, 385, 337

\bibitem[{Towler {et~al.}(2024)Towler, Kay, Schaye, Kugel, Schaller, Braspenning, Elbers, Frenk, Kwan, Salcido, {et~al.}}]{towler2024inferring}
Towler, I., Kay, S.~T., Schaye, J., {et~al.} 2024, Monthly Notices of the Royal Astronomical Society, 529, 2017

\bibitem[{Tozzi {et~al.}(2000)Tozzi, Scharf, \& Norman}]{tozzi2000detection}
Tozzi, P., Scharf, C., \& Norman, C. 2000, The Astrophysical Journal, 542, 106

\bibitem[{Tweed {et~al.}(2009)Tweed, Devriendt, Blaizot, Colombi, \& Slyz}]{tweed2009building}
Tweed, D., Devriendt, J., Blaizot, J., Colombi, S., \& Slyz, A. 2009, Astronomy \& Astrophysics, 506, 647

\bibitem[{Urban {et~al.}(2011)Urban, Werner, Simionescu, Allen, \& B{\"o}hringer}]{urban2011x}
Urban, O., Werner, N., Simionescu, A., Allen, S., \& B{\"o}hringer, H. 2011, Monthly Notices of the Royal Astronomical Society, 414, 2101

\bibitem[{Vikhlinin {et~al.}(1999)Vikhlinin, Forman, \& Jones}]{vikhlinin1999outer}
Vikhlinin, A., Forman, W., \& Jones, C. 1999, The Astrophysical Journal, 525, 47

\bibitem[{Vogelsberger {et~al.}(2020)Vogelsberger, Marinacci, Torrey, \& Puchwein}]{vogelsberger2020cosmological}
Vogelsberger, M., Marinacci, F., Torrey, P., \& Puchwein, E. 2020, Nature Reviews Physics, 2, 42

\bibitem[{Vurm {et~al.}(2023)Vurm, Nevalainen, Hong, Bah{\'e}, Dalla~Vecchia, \& Hein{\"a}m{\"a}ki}]{vurm2023cosmic}
Vurm, I., Nevalainen, J., Hong, S., {et~al.} 2023, Astronomy and Astrophysics, 673, A62

\bibitem[{Wicker {et~al.}(2023)Wicker, Douspis, Salvati, \& Aghanim}]{wicker2023constraining}
Wicker, R., Douspis, M., Salvati, L., \& Aghanim, N. 2023, Astronomy \& Astrophysics, 674, A48

\bibitem[{{XRISM Science Team}(2020)}]{xrism2020science}
{XRISM Science Team}. 2020, arXiv preprint arXiv:2003.04962

\bibitem[{Zhang {et~al.}(2023)Zhang, Adhikari, Costanzi, Frieman, Annis, \& Chang}]{zhang2023effect}
Zhang, Y., Adhikari, S., Costanzi, M., {et~al.} 2023, The Open Journal of Astrophysics, 6, 46

\bibitem[{Zhuravleva {et~al.}(2013)Zhuravleva, Churazov, Kravtsov, Lau, Nagai, \& Sunyaev}]{zhuravleva2013quantifying}
Zhuravleva, I., Churazov, E., Kravtsov, A., {et~al.} 2013, Monthly Notices of the Royal Astronomical Society, 428, 3274

\end{thebibliography}

\appendix

\section{Comparison of median and mean pressure radial profiles in the regions of collapsing and outflowing materials}\label{mean_vs_med_appendix}

\begin{figure}[h]
            \centering
            \includegraphics[trim=0 20 0 20, width=.5\textwidth]{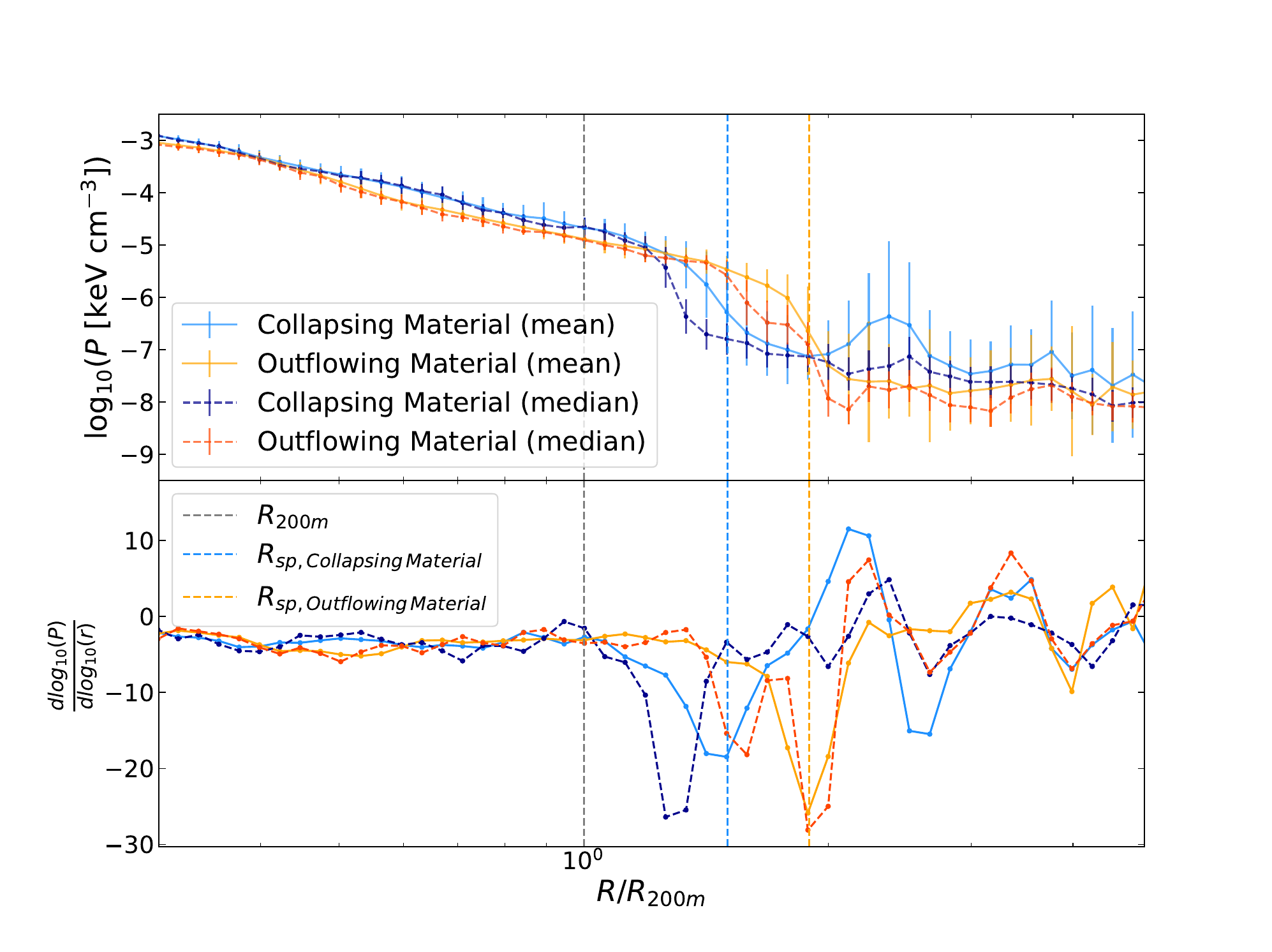}
            \caption{Pressure radial profiles in the regions of collapsing (solid blue for the mean and dashed dark blue for the median) and outflowing materials (solid orange for the mean and dashed dark orange for the median). The vertical dashed lines are $R_{\text{sp,\,Collapsing\,Material}}$ (blue), $R_{\text{sp,\,Outflowing\,Material}}$ (orange), and $R_{200m}$ (grey).} 
            \label{med_vs_mean_p}
\end{figure}

Fig. \ref{med_vs_mean_p} compares the mean (light profiles) and median (dark profiles) pressure radial profiles in the regions of collapsing (solid blue and dashed dark blue) and outflowing materials (solid orange and dashed dark orange). It shows that using the median instead of the mean to identify $R_{\text{sp}}$ gives the same result in the region of outflowing material. In the region of collapsing material, $R_{\text{sp}}$ is identified at a smaller radius in the median profile but still in agreement with that identified in the mean profile given the uncertainties. However, in the median profiles a steepening of the profile occurs at a smaller radius, although the median and mean profiles are less than 2$\sigma$ from each other at worst.

\section{2D-projected radial profiles of 100 stacked random projections compared to the mean and median profile of one projection} \label{stacking_appendix}

\begin{figure}
            \centering
            \includegraphics[trim=0 20 0 20, width=.41\textwidth]{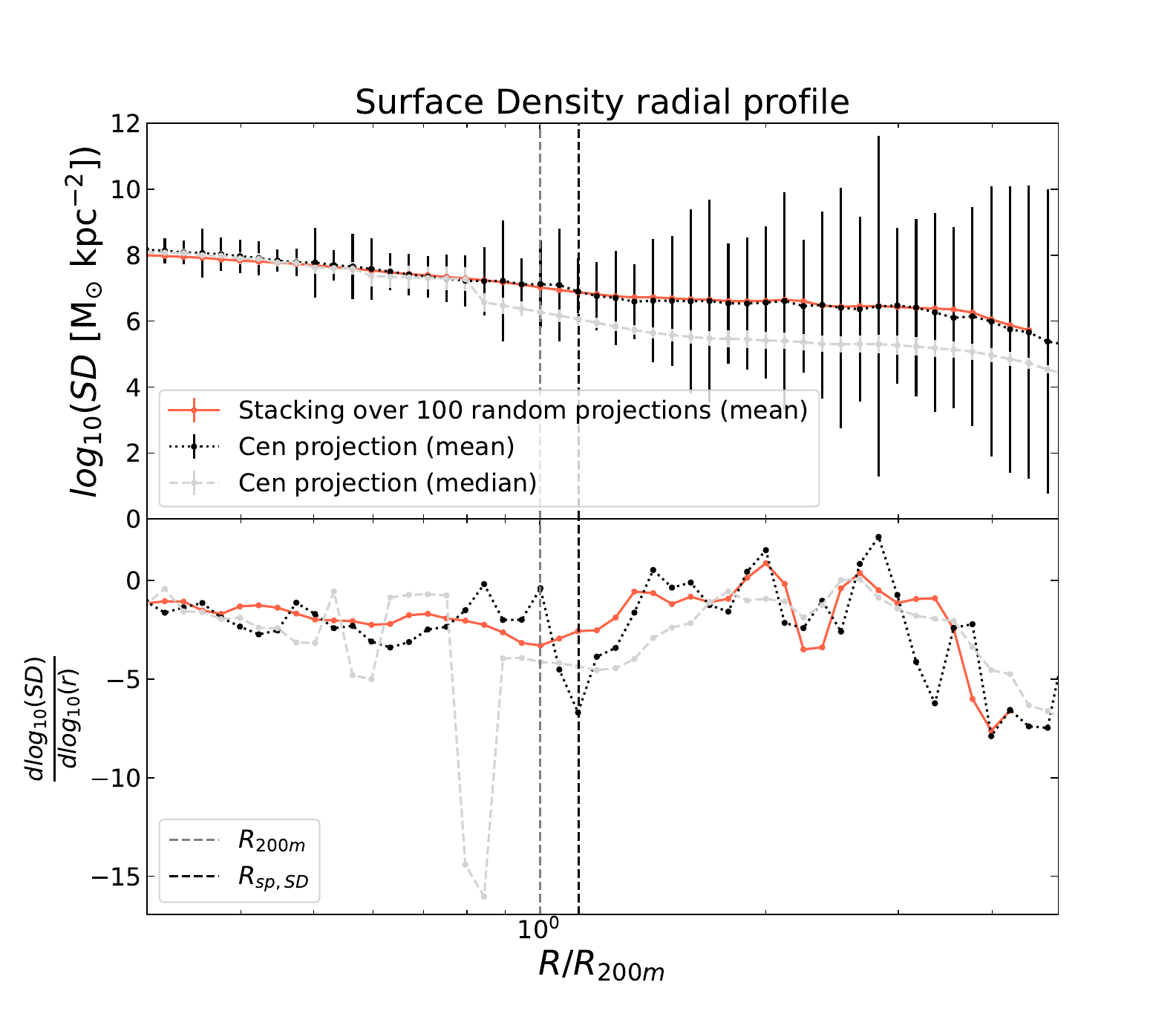}
            \caption{SD 2D-projected radial profile of 100 stacked random projections, displayed in solid light red. For comparison, we also show 2D-projected mean (dotted black) and median (dashed grey) radial profiles. The vertical dashed lines are $R_{\text{200m}}$ (grey) and $R_{\text{sp}}$ (black). }
            \label{stacked_profs_SD}
\end{figure}

\begin{figure}
            \centering
            \includegraphics[trim=0 20 0 20, width=.41\textwidth]{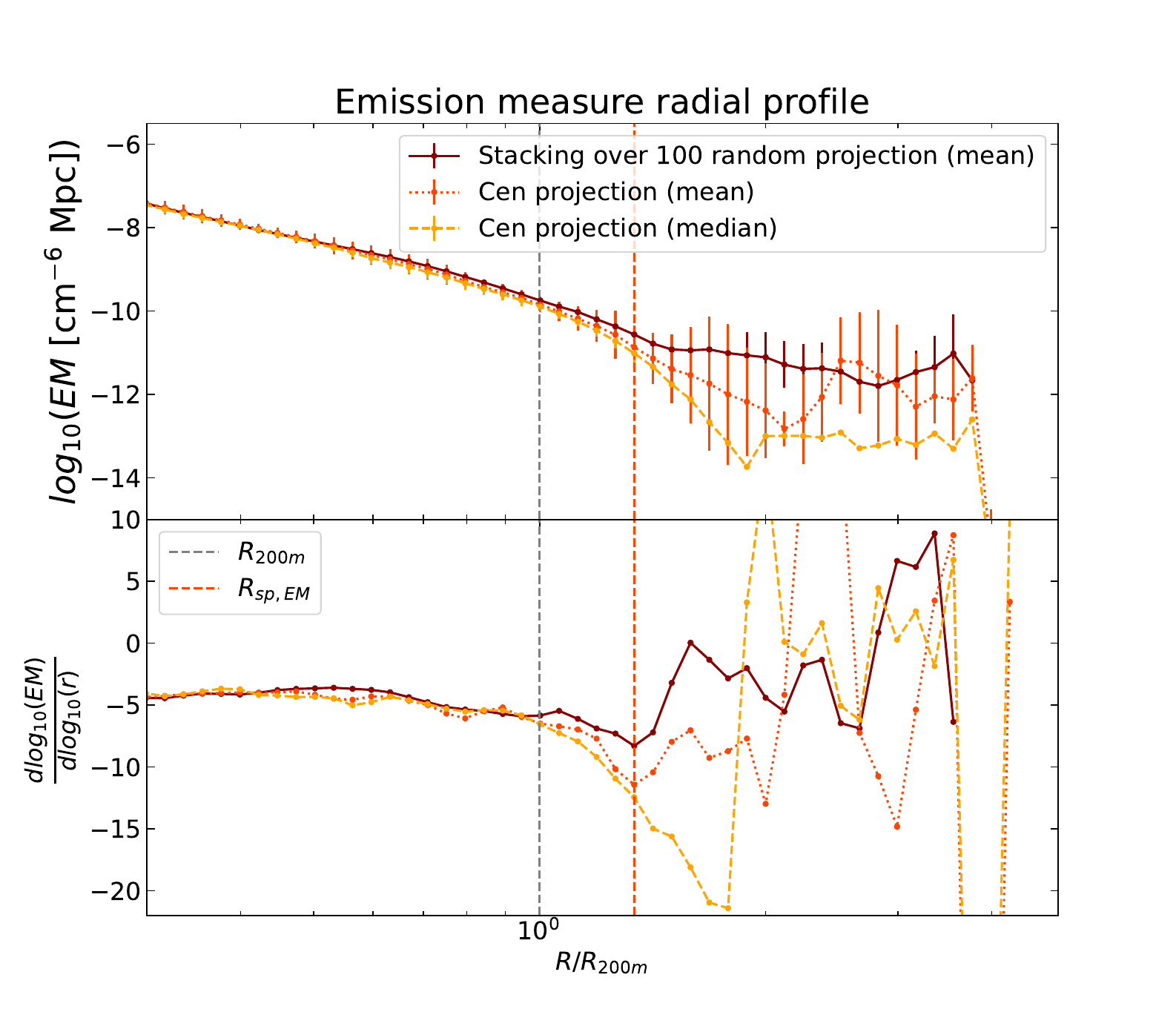}
            \caption{EM 2D-projected radial profile of 100 stacked random projections, displayed in solid dark red. For comparison, we also show 2D-projected mean (dotted red) and median (dashed orange) radial profiles. The vertical dashed lines are $R_{\text{200m}}$ (grey) and $R_{\text{sp}}$ (red). }
            \label{stacked_profs_EM}
            
\end{figure}

\begin{figure}
            \centering
            \includegraphics[trim=0 20 0 20, width=.41\textwidth]{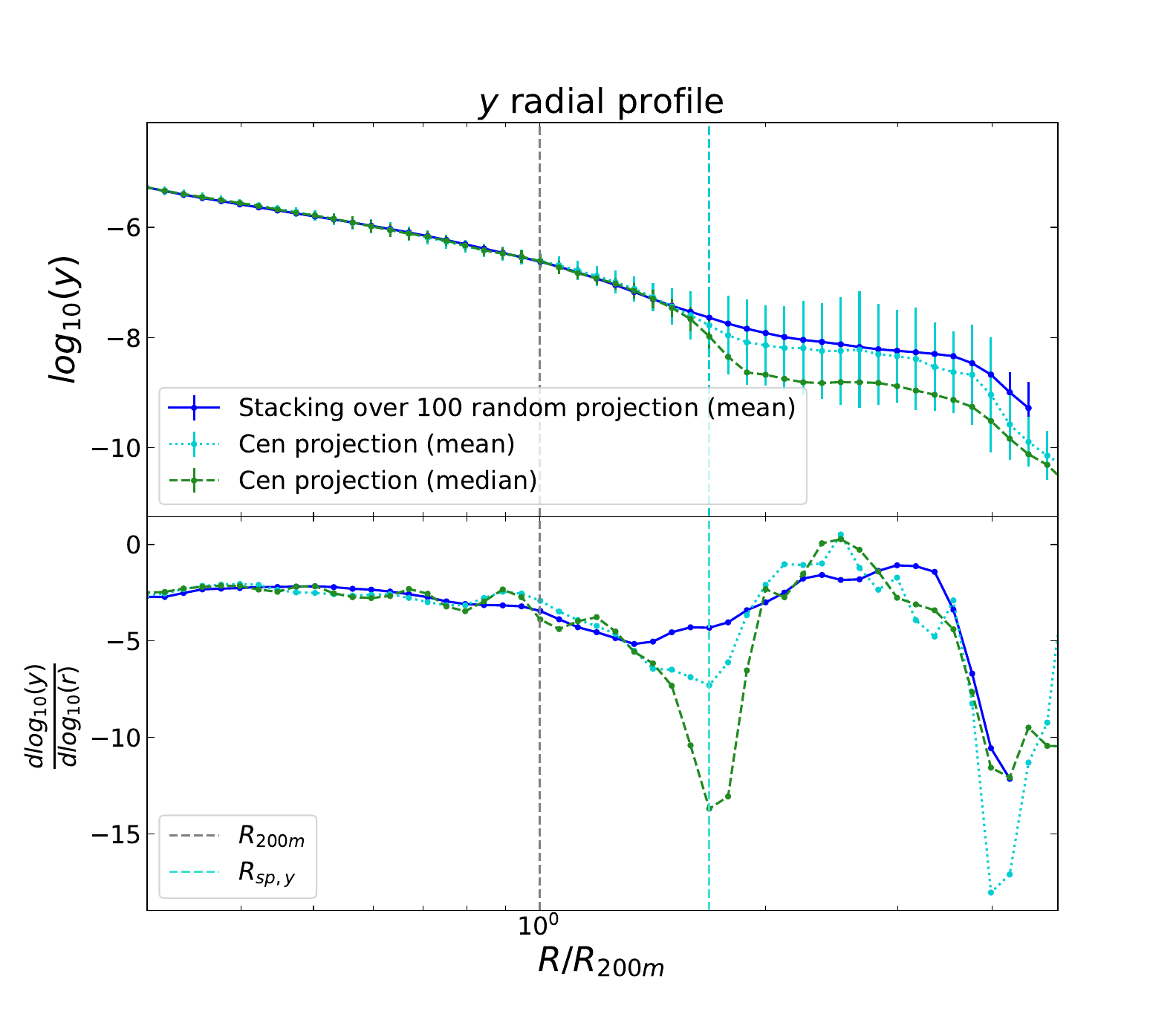}
            \caption{Compton-$y$ 2D-projected radial profiles of 100 stacked random projections, displayed in solid dark blue. For comparison, we also show 2D-projected mean (dotted cyan) and median (dashed light blue) radial profiles. The vertical dashed lines are $R_{\text{200m}}$ (grey) and $R_{\text{sp}}$ (cyan). }
            \label{stacked_profs_y}
            
\end{figure}

We present the SD (solid light red profile in Fig. \ref{stacked_profs_SD}), EM (solid dark red profile in Fig. \ref{stacked_profs_EM}), and Compton-$y$ (solid dark blue profile in Fig. \ref{stacked_profs_y}) 2D-projected radial profiles of 100 stacked random projections. The simulation box was rotated using randomly chosen angles to create 100 projections, from which we computed the maps and derived the 2D-projected mean profiles following the method presented in Sect.\ref{sec:2}. We then calculated the average profile over the hundred realisations. For comparison, we also show 2D-projected mean (dotted black, red, and cyan, respectively) and median (dashed grey, orange, and green, respectively) radial profiles. It shows that the stacked profile tends to have a higher signal in the outskirts than the mean or median profiles from the projection (named Cen in the figures) used in this work. This was expected because the filaments have various orientations in the random projections, leading to a smoothed stacked profile. We thus cannot identify $R_{\text{sp}}$ in any of these stacked 2D-projected profiles. We discuss the differences between the mean and median profiles in Sect.\ref{sec:5}.

\end{document}